\documentclass[a4paper,12pt]{article}
\pdfoutput=1
\AddToHook
{package/refstyle/before}
{\DeclareRobustCommand\eqref
{\relax}
\let\eqref\relax}

\usepackage{natbib}
\usepackage{epsfig,amsmath,amsfonts,amsthm}
\usepackage[left=28mm,right=20mm,top=30mm,bottom=20mm]{geometry}
\usepackage{subfig}
\usepackage{graphicx}
\usepackage{moreverb}

\DeclareGraphicsExtensions{.pdf,.png,.jpg}

\numberwithin{equation}{section}

\def\e{\hbox{E}}

\def\var{\hbox{Var}}

\def\min{\hbox{min}}
\def\max{\hbox{max}}

\usepackage[nokeyprefix]{refstyle}
\usepackage{varioref}
\usepackage{xr}
\usepackage{hyperref}
\usepackage{multirow}

\begin{document}

\title{Simulations for Meta-analysis of Magnitude Measures}

\author{Elena Kulinskaya and David C. Hoaglin }

\date{\today}

\maketitle

\begin{abstract}
Meta-analysis aims to combine effect measures from several studies. For continuous outcomes, the most popular effect measures use simple or standardized differences in sample means. However, a number of applications focus on the absolute values of these effect measures (i.e., unsigned magnitude effects). We provide statistical methods for meta-analysis of magnitude effects based on  standardized mean differences. We propose a suitable statistical model for random-effects meta-analysis of absolute standardized mean differences (ASMD), investigate a number of statistical methods for point and interval estimation, and provide practical recommendations for choosing among them.
\end{abstract}

\section{Introduction}
Meta-analysis aims to combine effect measures from several studies. For continuous outcomes, the most popular effect measures use simple or standardized differences in sample means. However, a number of applications focus on the corresponding magnitudes, without regard to their direction.

Meta-analyses of magnitude effects are quite common in ecology and evolutionary biology, in situations where the direction of the effect is less important.  As a rationale, \cite{Garamszegi2006} argued that \lq\lq the mean of the absolute values of the effect sizes may show that weak or strong effects are at work in general without considering directional roles'' or \lq\lq the researcher may want to compare unsigned effect sizes between different groups of traits, such as between plumage and song traits.'' \cite{Clements2022} studied the impacts of ocean acidification on fish behavior and used the absolute value \lq\lq due to the inherent difficulty in assigning a functional direction to a change in behavior, as many behavioral changes can be characterized by both positive and negative functional trade-offs". \cite{Koricheva2023} studied  physical and chemical leaf traits that could affect herbivory but \lq\lq expected the direction of the effect to be highly context-dependent (i.e., different neighbours may cause either an increase or a decrease in the same leaf trait)''.
Other examples include \cite{Bailey2009} (absolute effects of plant genetic factors across levels of organization), \cite{Champagne2016} (influence of the neighboring plant on the focal plant herbivory level), and \cite{Costantini2017} (sexual differentiation in resistance to oxidative stress  across vertebrates).

\cite{morrisey-2016} discussed the rationale for magnitude effects in evolutionary biology and proposed some statistical methods for meta-analysis of absolute mean values. We discuss his work in Section 2.1. However, the majority of the cited papers used the absolute standardized mean difference (ASMD), though some used the absolute values of Pearson correlation or log-response ratio. Interestingly, ASMD values are routinely used for testing the balance of individual  covariates between the two groups of an observational study when assessing the quality of a propensity-scores-based model, with 0.1 as the standard cutoff \citep{rubin-2001, Sanni-2019}.

Typically, the systematic reviews  include meta-analyses of both directional and unsigned effects. Worryingly, to meta-analyze their absolute values (magnitude effects), those reviews (\cite{Champagne2016, Costantini2017, Clements2022, Koricheva2023}) use routine inverse-variance methods developed for directional effects, which have very different statistical properties.
The likely explanation is the lack of  statistical methods specifically for MA of magnitude effects. This article aims to fill this important gap.
We develop  statistical methods for meta-analysis of  ASMD-based magnitude effects and study their performance by simulation.

\section{Notation}

We assume that each of the $K$ studies in the meta-analysis consists of two arms, Treatment and Control, with sample sizes $n_{iT}$ and $n_{iC}$. The total sample size in Study $i$ is $n_i = n_{iT} + n_{iC}$. We denote the ratio of the Control sample size to the total by  $f_i = n_{iC} / n_{i}$.  The subject-level data in each arm are assumed to be normally distributed with means $\mu_{iT}$ and $\mu_{iC}$ and variances $\sigma_{iT}^2$ and $\sigma_{iC}^2$. (We appreciate, however, that real data are not exactly normal.) The sample means are $\bar{x}_{ij}$, and the sample variances are $s^2_{ij}$, for $i = 1, \ldots, K$ and $j = C$ or $T$.

\section{Absolute mean difference}
The mean difference (MD) effect measure  is
\[ \mu_{i} = \mu_{iT} - \mu_{iC}, \hbox{    estimated by    } y_i = \bar{x}_{iT} - \bar{x}_{iC}, \]
with variance
$\sigma^2_i = \sigma_{iT}^2 / n_{iT} + \sigma_{iC}^2 / n_{iC}$, estimated by
\begin{equation} \label{eq:varMD}
v_i^2 = \hat{\sigma}_i^2 = s_{iT}^2 / n_{iT} + s_{iC}^2 / n_{iC}.
\end{equation}
%The sample variances $s_{iT}^2$ and $s_{iC}^2$ do not depend on $\mu_{iT}$ and $\mu_{iC}$, so $\hat{\sigma}_i^2$ does not involve $\mu_i$. %In the best-case scenario for traditional meta-analysis methods,
%(For normal data, the sample means are independent of the sample variances.) % (and therefore of inverse-variance-based weights).
%However, the relation of the between-study variance $\tau^2$ and the within-study variances $\sigma_i^2$  may affect quality of estimation.
Sometimes the  pooled sample  variance is used instead of $v_i^2$. Then, however, unequal variances in the Treatment and Control arms can adversely affect estimation \citep{kulinskaya2004welch}.

The familiar common-effect model for MD assumes that $\mu_i = \mu$ for all $i$, whereas the random-effects model allows the $\mu_i$ to come from a distribution with mean $\mu$ and variance $\tau^2$, usually $N(\mu, \tau^2)$. Point estimation of $\mu$ often uses a weighted mean, $\hat{\mu} = (\Sigma w_i y_i) / (\Sigma w_i)$, with $w_i = 1 / \hat{\sigma}_i^2$ in the common-effect model and $w_i = 1 / (\hat{\sigma}_i^2 + \hat{\tau}^2)$ in the random-effects model. Several popular methods base estimators of $\tau^2$ on $Q = \Sigma w_i (y_i - \bar{y}_w)^2$, with $\bar{y}_w = (\Sigma w_i y_i) / (\Sigma w_i)$ and, initially, $w_i = 1 / \hat{\sigma}_i^2$. We return to these methods in Section~\ref{sec:simest}. %Appendix~\ref{sec:REMandQ} contains further details.

The underlying normal distributions in the two arms result in normality of MD: $y_i \sim N(\mu_i, \sigma^2_i)$. Hence, the absolute mean difference (AMD) $|y_i|$ has a folded normal distribution $FN(\mu, \sigma^2)$ (\cite{Leone1961}, \cite[p.453]{J-K-B-1995}, \cite{tsagris-2014}).   For simplicity of notation, we sometimes drop the subscript $i$.

The first two moments of the $FN(\mu, \sigma^2)$ distribution are
\begin{equation} \label{eq:momAMD}
\mu_f = \e(|y|) = 2 \sigma \phi(\mu / \sigma) + \mu \left[ 1 - 2 \Phi(- \mu / \sigma) \right],\;\; \sigma^2_f = \mu^2 + \sigma^2 - \mu_f^2,\end{equation}
where $\phi(\cdot)$ and $\Phi(\cdot)$ are the density and the cdf of the standard normal distribution. \cite{tsagris-2014} give the moment-generating function and higher moments and the maximum-likelihood estimators of the parameters. When $\mu = 0$, $FN(\mu, \sigma^2)$ is a half-normal distribution with mean $\sigma (2 / \pi)^{1/2}$ and
variance $\sigma^2 (1 - (2 / \pi) )$.  A  difference $|y_i|-v_i(2/\pi)^{1/2}$ could be used as a centered-at-zero absolute mean effect measure, as suggested in \cite{morrisey-2016}.

From  Equation~(\ref{eq:momAMD}), the expected $| y_i |$ depends on both the standardized mean $\delta_i = \mu_i / \sigma_i$ and the variance $\sigma_i^2$, so AMD does not seem to be an appropriate effect measure for magnitude.
Additionally,
its variance is rather difficult to estimate. A na\"{i}ve estimate would be $\hat \sigma^2_f = y^2 + v^2 - \hat\mu_f^2$. Substituting the MD $y$ and its standard deviation $v$ in the expression for $\mu_f$ in Equation~(\ref{eq:momAMD}) results in an $O(1/n)$ biased estimate of $\mu_f$ and, therefore, of its variance. It is possible to eliminate this bias by using the second-order Taylor expansion of $h(\mu,\sigma)=\mu_f$, but the corrected estimate appears to be rather complicated.
%To eliminate this bias, we write out  the 2nd order Taylor expansion of $h(\mu,\sigma)=\mu_f$ and use the unbiasedness of $y$ and $v$ to obtain:
%$$\e(h(y, v))= h(\mu, \sigma)+\frac{\var(y)}{2}h''_{\mu\mu}(\mu, \sigma)+\frac{\var(v)}{2}h''_{\sigma\sigma}(\mu, \sigma).$$
%A corrected estimate of $\mu_f$ appears to be rather complicated.
%$$\hat\mu_f=h(y, v)-\frac{s^2_n}{2n}h''_{\mu\mu}(\bar x, s_n)-\frac{s_n^2}{4n}h''_{\sigma\sigma}(\bar x, s_n).$$
%to Corrected  estimate of $\mu_f$ is obtained by

To summarize, dependence on the nuisance parameter $\sigma_i^2$, lack of asymptotic normality, and difficulty in estimating the variance of AMD preclude use of AMD in meta-analysis.
Dividing $\mu_f$ in Equation~(\ref{eq:momAMD}) by $\sigma$ results in a simpler expression that depends on only the standardized mean $\delta = \mu / \sigma$ and appears to be much more convenient for further analysis, suggesting use of ASMD instead. Therefore, we abandon AMD in favor of ASMD in what follows.

\section{Absolute standardized mean difference} \label{sec:EffectSMD}

The standardized mean difference effect measure is
\[ \delta_{i}=\frac{\mu_{iT}-\mu_{iC}}{\sigma_{i}}. \]
The variances in the Treatment and Control arms are usually assumed to be equal. Therefore, $\sigma_i$ is estimated by the square root of the pooled sample variance
\begin{equation}\label{eq:pooledvar}
s_i^2=\frac{(n_{iT}-1)s_{iT}^2 +(n_{iC}-1)s_{iC}^2}{n_{iT}+n_{iC}-2}.
\end{equation}
The plug-in estimator $d_i = (\bar{x}_{iT} - \bar{x}_{iC}) / s_{i}$, known as Cohen's $d$, is biased in small samples. \cite{hedges1983random} derived the unbiased estimator
$${g}_i = J(m_i) \frac{\bar{x}_{iT} - \bar{x}_{iC}} {s_{i}},$$
where $m_{i} = n_{iT} + n_{iC} - 2$, and $$J(m) = \frac{\Gamma \left(\frac{m} {2}\right)} {\sqrt{\frac{m} {2}}\Gamma \left( \frac{m - 1} {2} \right)},$$
often approximated by $1 - 3 / (4m - 1)$. This estimator of $\delta_i$, typically used in meta-analysis of SMD,  is sometimes called Hedges's $g$.
%The unbiased estimator for the variance of ${g}_i$ is \citep{hedges1983random}
%\begin{equation}\label{eq:g_var}
%v_{i}^2=\frac{n_{iT}+n_{iC}}{n_{iT}n_{iC}}+\left(1-\frac{(m_{i}-2)}
%{m_{i}J(m_{i})^2}\right)g^2_{i}.
%\end{equation}

Denote by $\tilde n_i = n_{iC} n_{iT} / n_i = n_i q_i (1 - q_i)$ the effective sample size in Study $i$.  The sample SMD $d_i$ (and therefore Hedges's estimate ${g}_i$) has a scaled noncentral $t$-distribution with noncentrality parameter (NCP) $\tilde n_i^{1/2} \delta_i$:
\begin{equation}\label{eq:g_dist}
\tilde n_i^{1/2} {d}_i \sim t_{m_i}(\tilde n_i^{1/2} \delta_i).
\end{equation}
Therefore, the ASMD $|d_i|$ has  a {\it folded} scaled noncentral $t$-distribution with the same noncentrality parameter:
\begin{equation}\label{eq:g_A_dist}
\tilde n_i^{1/2} {|d_i|} \sim FNT_{m_i}(\tilde n_i^{1/2} \delta_i).
\end{equation}
Alternatively, $d_i^2$ has a scaled noncentral $F_{1,m_i}(\tilde n_i \delta_i^2)$ distribution.

A central folded $t$-distribution has $\mu = 0$, and a half-$t$ additionally has $\sigma = 1$. The half-$t$ was introduced  by \cite{psarakis-1990}, who derived its moments and discussed its relations to other distributions. In particular, when $\nu \to \infty$, the folded $t_\nu$ converges to the folded normal distribution.

\cite{Gelman-2006} introduced the FNT distribution as a noninformative conditionally-conjugate prior for the standard deviation $\tau$  of the variance component in random-effects meta-analysis. % It was subsequently used in this way in \cite{thompson-2020} and \cite{rover-2021}.
However, we have not found any publications on the moments of the FNT distribution.

\section{Squared standardized mean difference} \label{sec:EffectSMD2}

The square of a FNT$(\lambda)$ random variable with $\nu$ df  has a non-central $F_{1,\nu}(\lambda^2)$-distribution, as does the square of a noncentral $t$ random variable. As $\nu \to \infty$, the distribution $F_{1,\nu}(\lambda^2)$ converges to the noncentral $\chi^2_1(\lambda^2)$. And when $\lambda^2 \to 0$, the distribution converges to the central $F_{1,\nu}$ distribution .

The first and second moments of the
noncentral $F(\lambda^2)$ distribution (the special case of the doubly-noncentral $F$-distribution $F_{\nu_1, \nu_2}(\lambda_1, \lambda_2)$ with $\lambda_1 = \lambda^2$ and $\lambda_2 = 0$) with $\nu_1,\; \nu_2 > 4$ are \cite[(30.3)]{J-K-B-1995}
\begin{equation}\label{eq:F_mom}
\e(X)=\frac{\nu_2 (\nu_1 + \lambda^2)}{\nu_1 (\nu_2 - 2)},\;\;\; \var(X)=2 \left(\frac{\nu_2}{\nu_1} \right)^2 \frac{(\nu_1 + \lambda^2)^2 + (\nu_1 + 2 \lambda^2) (\nu_2 - 2)} {(\nu_2 - 2)^2 (\nu_2 - 4)}.
\end{equation}
From Equation~(\ref{eq:g_dist}),   %the distribution of $g_i$ from
$$d_i^2 \sim \tilde n_i^{-1} F_{1,m_i}(\tilde n_i \delta_i^2).$$
Using  $\nu_{1} = 1$ and $\nu_{2} = m_i$ in Equation~(\ref{eq:F_mom}), the  moments of  $d_i^2$ are
\begin{equation}\label{E_g_i}
\e(d_i^2) = \left( \frac{m_i} {m_i - 2} \right) (\tilde n_i^{-1} + \delta_i^2),
\end{equation}

\begin{equation}\label{var_g_i}
\var(d_i^2) = \frac{2 m_i^2} {(m_i - 2)^2 (m_i - 4)} \left( \frac{m_i - 1} {\tilde n_i^2} + \frac{2 (m_i - 1) \delta_i^2}{\tilde n_i} + \delta_i^4 \right).
\end{equation}

From Equation~(\ref{E_g_i}), an unbiased estimate of the squared SMD $\delta^2$ is
\begin{equation}\label{hat_delta2} \widehat{\delta^2_i} = \frac{m_i - 2} {m_i} d_i^2 - \frac{1} {\tilde n_i}.\end{equation}
% where $d_i$ is the Cohen's $d$ and $\tilde n_i d_i^2\sim F_{1,m_i}(\tilde n_i \delta_i^2)$.%, where $\tilde n_i= n_iq_i(1-q_i)$ is the effective sample %size.
The variance of $\widehat{\delta^2_i}$ is
\begin{equation}\label{var_delta2}
\var(\widehat{\delta^2_i}) = \frac{2} {(m_i - 4)} \left( \frac{m_i - 1} {\tilde n_i^2} + \frac{2 (m_i - 1) \delta_i^2} {\tilde n_i} + \delta_i^4 \right).
\end{equation}
Combining Equations~(\ref{hat_delta2}) and (\ref{var_delta2}),
$$\e(d_i^4) = \frac{m_i^2} {(m_i - 2) (m_i - 4)} \left( \frac{3} {\tilde{n}_i^2} + 6 \frac{\delta_i^2} {\tilde{n}_i} + \delta_i^4 \right).$$
Hence,
$$\widehat{\delta_i^4} = \frac{(m_i - 2)(m_i - 4)} {m_i^2} d_i^4 - \frac{6} {\tilde{n}_i} \frac{m_i - 2} {m_i} d_i^2 + \frac{3} {\tilde{n}_i^2}.$$
Substituting $\widehat{\delta^2_i}$ from Equation~(\ref{hat_delta2}) and the above estimate of $\widehat{\delta^4_i}$ into Equation~(\ref{var_delta2}), we  obtain an unbiased estimate of $\var(\widehat{\delta^2_i})$ :
\begin{equation}\label{hatvar_delta2}
\widehat{\var}(\widehat{\delta^2_i}) =
%\frac{2}{(m_i-2)}\big(\frac{m_i-1}{\tilde n_i^2}+\frac{2(m_i-1)\widehat{\delta_i^2}}{\tilde n_i}+\widehat{\delta_i^4}\big)\=
\frac{2 (m_i - 2)} {m_i^2} d_i^4 + \frac{4 (m_i - 2)} {m_i \tilde n_i} d_i^2 - \frac{2} {\tilde n_i^2}  .
\end{equation}

The related problem of estimating the noncentrality $\lambda^2$ from a single observation $F'$ from $F_{\nu_1,\nu_2}(\lambda^2)$ is well investigated. The UMVUE estimator is $\hat \lambda^2 = \nu_1 \nu_2^{-1} (\nu_2 - 2) F' - \nu_1$, which, for our setting, becomes $\widehat{\delta^2_i}$ but is inadmissible, as is its truncated-at-zero version. See \cite[Section 30.6]{J-K-B-1995} for discussion of point and interval estimation of $\lambda^2$.

\cite{steiger2004} provides an explicit algorithm for finding a $(1 - \alpha)$ confidence interval for the noncentrality parameter of a noncentral $F$ distribution $F(\cdot;\lambda^2)$  based on an inverted $F$ test. We obtain a confidence interval for $\delta_i^2$ as follows:
\begin{itemize}
\item  Calculate $1 - p = F_{1,m_i}(\tilde n_i d_i^2;0)$.
\item  If $1 - p < \alpha / 2$, $\lambda^2_{upper} = 0$. Otherwise, solve for $\lambda^2_{upper}$ in $F_{1,m_i}(\tilde n_i d_i^2;\lambda^2_{upper}) = \alpha/2$.
\item  If $1 - p < 1 - \alpha/2)$, $\lambda^2_{lower} = 0$. Otherwise, solve for $\lambda^2_{lower}$ in $F_{1,m_i}(\tilde n_i d_i^2;\lambda^2_{lower}) = 1 - \alpha/2$.
\item The confidence interval for $\delta_i^2$ is $\tilde n_i^{-1} (\hat\lambda^2_{lower},\;\hat\lambda^2_{upper})$, and taking the square root of these estimated confidence limits yields the confidence interval for $|\delta|$.
\end{itemize}
The above equations for the confidence limits have a unique solution because $F_{\nu_1,\nu_2}(\cdot;\lambda^2)$ is a decreasing function of $\lambda^2$.
We call these confidence intervals, based on inverted $F$ or $\chi^2$ tests, $F$- or $\chi^2$-profile intervals.

\section {Meta-analysis of squared SMD}

We assume that the $K$ studies, with sample sizes $(n_{iC}, n_{iT})$ in the Control and Treatment arms, respectively,  resulted in
magnitude effects $d_i^2$ or $\widehat{\delta_i^2},\; i = 1,\ldots, K$. We formulate  common-effect  and random effects models (REM) for magnitude effects in sections~\ref{sec:FEM} and \ref{sec:REM}, respectively. Inference for $\delta^2$ under REM is discussed in sections~\ref{sec:delta2_orig} and \ref{sec:cond}.

\subsection{Common-effect model for $\delta^2$}\label{sec:FEM}

We formulate the common-effect model (also known as the fixed-effect model) for the magnitude effect as
\begin{equation}\label{eq:FEM}
\tilde{n}_i {d_i^2} \sim F_{1,m_i}(\tilde{n}_i \delta^2) ,\;\; i = 1,\ldots, K.
\end{equation}
The objective is to estimate the magnitude $\delta^2$.

From Equation~(\ref{hat_delta2}), any weighted average of the $\widehat{\delta_i^2}$ is an unbiased estimate of $\delta^2$. The simplest choice uses weights proportional to $\tilde n_i$. Then
\begin{equation}\label{eq:Delta_FEM}
\widehat\delta^2 = (\Sigma \tilde n_i)^{-1} \sum_1^K \tilde{n}_i \widehat{\delta_i^2} = (\Sigma \tilde n_i)^{-1} \left[ \sum_1^K \frac{m_i - 2}{m_i} \tilde{n}_i d_i^2 - K \right]
\end{equation}
is distributed as a shifted and scaled sum of $F_{1,m_i}(\tilde{n}_i \delta^2)$-distributed r.v.'s. Also, the simpler statistic
\begin{equation}\label{eq:CDF_FEM}
d^2 = (\Sigma \tilde n_i)^{-1} \Sigma \tilde n_i d_i^2 \sim (\Sigma \tilde n_i)^{-1} \left[ \sum_1^K F_{1,m_i}(\tilde n_i \delta^2) \right].
\end{equation}
%\begin{equation}\label{eq:Delta_FEM}
%\widehat\delta^2\sim (\sum \tilde n_i)^{-1}\big[\sum_1^K\frac{m_i-2}{m_i}F_{1,m_i}(\tilde n_i\delta^2)-K\big].
%\end{equation}
This distribution appears rather complicated, and we are not aware of any publications or implementations of it.
  %A sum of central F r.v.'s was studied by \cite{Du2020} who provided  an exact analytical expression to its density and a three-moment approximation by a % single F distribution.
When $\tilde{n}_i \to \infty$, it converges to a scaled (by $(\sum \tilde n_i)^{-1}$) sum of  $\chi^2_1(\tilde{n}_i \delta^2)$ distributions, which is just a scaled noncentral $\chi^2_K (\delta^2 \Sigma \tilde n_i)$ distribution \cite[(29.5)]{J-K-B-1995}:
\begin{equation}\label{eq:ACDF_FEM}
d^2 = (\Sigma \tilde n_i)^{-1} \sum \tilde n_i d_i^2 \underset{\{m_i\} \to \infty} {\sim} (\Sigma \tilde n_i)^{-1} \chi^2_K(\delta^2 \Sigma \tilde n_i).
\end{equation}
The statistic $(\sum \tilde n_i) d^2$ can be used to test for $\delta^2 = 0$ using the percentage points of the central $\chi^2_K$ distribution, in the case of large sample sizes,  or of the  central version of Equation~(\ref{eq:CDF_FEM}) directly by using the parametric bootstrap. % \citep{Du2020}.
 An algorithm similar to that at the end of Section~\ref{sec:EffectSMD2} can be used to obtain an approximate $(1 - \alpha)$-level $\chi^2$-profile  confidence interval for $\delta^2$. % with the following changes: use $\tilde n=\sum\tilde n_i$ instead of $\tilde n_i$ and  $\chi^2_K(\delta^2\sum\tilde n_i)$ instead of %$F_{1,m_i}(\tilde n_i\delta^2)$.

\subsection{Random-effects model for $\delta^2$}\label{sec:REM}

We formulate the random-effects model for the magnitude effect as
\begin{equation}\label{eq:REM}
\tilde{n}_i {d_i^2} \sim F_{1,m_i}(\tilde{n}_i \delta_i^2) ,\;\; \delta_i \sim N(\delta, \tau^2),\;\; i = 1,\ldots, K.
\end{equation}
The model for the $\delta_i$ is the standard random-effects model, with parameters $\delta$ and $\tau^2$. The objective, however, is to estimate $\delta^2$ instead of $\delta$. From $\delta_i / \tau \sim N(\delta / \tau, 1)$ we obtain $\delta_i^2 \sim \tau^2 \chi^2_1(\delta^2 / \tau^2)$.

The distribution of $\tilde{n}_i {d_i^2}$ in Equation~(\ref{eq:REM}) is conditional on $\delta_i^2$. Taking into account the distribution of $\delta_i$, $\tilde{n}_i {d_i^2}$ has a noncentral $F$-distribution mixed over its noncentrality parameter. By definition, the doubly-noncentral $F$-distribution $F(p, q, \lambda_1, \lambda_2)$ is the distribution of the ratio of two independent noncentral chi-square random variables: $F(p, q, \lambda_1, \lambda_2) = q X_1 / p X_2$, where $X_1 \sim \chi^2_p(\lambda_1)$ and $X_2 \sim \chi^2_q(\lambda_2)$. Corollary 2 of \cite{jones2021} states that if $F|(Y_1 = y_1, Y_2 = y_2) \sim F(p, q, h_1 y_1, h_2 y_2)$ and $Y_1 \sim \chi^2_p(\lambda_1)$ and $Y_2 \sim \chi^2_q(\lambda_2)$ independently, then $(1 + h_2) F / (1 + h_1) \sim  F(p, q, \frac{h_1 \lambda_1} {1 + h_1}, \frac{h_2 \lambda_2} {1+h_2} )$.

For $\tau^2>0$, we take $h_2 = 0$, $p = 1$, $q = m_i$, $h_1 = \tilde n_i\tau^2$, and $\lambda_1 = \delta_i^2 / \tau^2$ and write $\delta_i^2 / \tau^2 \sim \chi^2_1(\delta^2 / \tau^2)$ to obtain
\begin{equation}\label{eq:REMmarg}
\tilde{n}_i d_i^2 \sim (1 + \tilde n_i \tau^2) F_{1, m_i} \left( \frac{\tilde{n}_i \delta^2}{1 + \tilde n_i \tau^2} \right),\;\; i=1,\ldots, K.
\end{equation}
When $\tau^2 = 0$, Equation~(\ref{eq:REMmarg}) is still valid and  reduces to Equation~(\ref{eq:FEM}); that is, the random-effects model becomes the common-effect model.
Under the REM, %this estimator of $\delta^2$ is biased:
$$\e(\tilde{n}_i d_i^2) = \frac{m_i} {m_i - 2} (1 + \tilde n_i \tau^2 + \tilde n_i \delta^2) \text{   and   } \e(\widehat{\delta}_i^2) = \tau^2 + \delta^2$$.
Therefore,  $\widehat{\delta}^2$ given by Equation~(\ref{eq:Delta_FEM}) or any other weighted mean of the $\widehat{\delta_i^2}$ with constant weights would provide an unbiased estimate of $\tau^2 + \delta^2$.

%Also,
%\begin{equation}\label{eq:Delta_REM}
%\widehat\delta^2\sim (\sum \tilde n_i)^{-1}\big[\sum_1^K\frac{m_i-2}{m_i}(1+\tilde n_i\tau^2)F_{1,m_i}(\frac{\tilde n_i\delta^2}{1+\tilde %n_i\tau^2})-K\big].
%\end{equation}
%Unfortunately, it does not seem to be possible to estimate $\delta^2$ and $\tau^2$ separately, but only their sum, or  their ratio $\delta^2/\tau^2$.
%When $m_i\to\infty$,
%$$\widehat\delta^2\sim (\sum \tilde n_i)^{-1}\big[(1+\tau^2)\chi^2_K)(\frac{\tilde n_i\delta^2}{1+\tilde n_i\tau^2})-K\big].$$

\subsection{Inference for $\delta^2$ from signed values of SMD} \label{sec:delta2_orig}

When the initial meta-analysis used the $\hat\delta_i$ and estimated $\tau^2$ by $\hat \tau^2$, we can obtain a point estimate of the magnitude effect $\delta^2$ as
$\widehat{\widehat{\delta^2}} = \widehat{\delta^2} - \hat \tau^2$ or its truncated-at-zero version.

It is convenient to consider using a level $(1 - \alpha)$ confidence interval for $\delta$, $(L, U)$, as the basis for a level $(1 - \alpha)$ confidence interval for $\delta^2$.

%Let, without loss of generality, $0<L<U$. The na\"ive CI $(L^2, U^2)$ has level $> 1 - \alpha$ because $-U < \delta < -L$ and $L < \delta < U$ both yield $L^2 < %\delta^2 < U^2$. Section~\ref{sec:ApxA1} examines the details for $0 < L < U$ and $L < U < 0$ and, separately, for $L < 0 <U$, which leads to the na\"ive CI %$[0, \max(L^2, U^2)]$  for $\delta^2$. Appendix~\ref{sec:ApxA1} also derives corrections to obtain  confidence intervals at a confidence level closer to %nominal.

%\subsection{Na\"ive and corrected confidence intervals for $\delta^2$} \label{sec:ApxA1}

%Section~\ref{sec:delta2_orig} discusses  confidence intervals for $\delta^2$ based on  the confidence intervals for $\delta$ from the initial meta-analysis.
 By $I_{1-\alpha}(\delta)=(L,U)$ we denote a level-$(1-\alpha)$ CI for $\delta$ with level $(1 - \alpha)$. To allow unequal division of $\alpha$ between the two tails, we let $\beta < \alpha$ be the part in the  upper  tail. $L = \hat\delta - c_{1 - \beta} v(\hat\delta)$, $U = \hat\delta - c_{\alpha-\beta} v(\hat\delta)$, $v(\hat\delta)$ is the estimated standard deviation of $\hat\delta$, and $c_{\gamma}$ is the critical value at tail area $\gamma$ from an appropriate symmetric distribution $G$, such as normal or t.

%As discussed in Section~\ref{sec:delta2_orig},
When both confidence limits are on the same side of zero, say $0 < L < U$ (i.e., when $\hat\delta / v(\hat\delta) > c_{1-\beta}$), the na\"ive CI $(L^2,U^2)$ provides a  CI for $\delta^2$ with level $(1 - \gamma) \geq (1 - \alpha)$ for some $0 < \gamma < \alpha$ because $(L^2, U^2)$ also includes the values of $\delta$ in $-U < \delta < -L$. This extra coverage probability is
 \begin{equation} \label{eq:extraLU}
 \begin{array}{ll}
 P(-U < \delta < -L) & = P(-\hat\delta + c_{\alpha - \beta} v(\hat\delta) < \delta < -\hat\delta + c_{1 - \beta}v(\hat\delta))\\
                               & = P(c_{\alpha - \beta} < (\hat\delta - \delta + 2 \delta) / v(\hat\delta) < c_{1 - \beta})\\
                               & = G(c_{1 - \beta} - 2 \delta / v(\hat\delta)) - G(c_{\alpha - \beta} - 2 \delta / v(\hat\delta)).\end{array}
\end{equation}
%Equation~(\ref{eq:extraLU}) also holds when $L < U <0$.

When $\beta = \alpha / 2 = .025$, the probability $P(-U < \delta < -L)$ decreases from
 .025 when $\delta / v(\hat\delta) = c_{1 - \beta}$ to 4.43e-05 when $\delta / v(\hat\delta) = 3 c_{1 - \beta} / 2$ to 2.052e-09  when $\delta / v(\hat\delta) = 2 c_{1 - \beta}$. The case $L < U <0$ yields the same values when  $-\hat\delta / v(\hat\delta) > c_{\alpha - \beta}$. The extra coverage seems small enough not to require correction of the confidence level.

However, to obtain exactly  level $(1-\gamma)$ coverage for $\delta^2$ for an arbitrary $\gamma$, take, for simplicity, $\beta = \alpha / 2$, substitute $\hat\delta$ for $\delta$ in Equation~(\ref{eq:extraLU}), and solve for $\alpha$ in the equation $\gamma = \alpha - \hat P(-U < \delta < -L)$.

Similarly, when $L < 0 <U$ or, equivalently, when $c_{\alpha - \beta} < \hat\delta / v(\hat\delta) < c_{1-\beta}$,   we can choose  the na\"ive confidence  interval  $I_{1 - \gamma} (\delta^2) = [0, \max(L^2,U^2))$ for $\delta^2$. This interval provides a CI for $\delta^2$ with level $(1 - \gamma)\geq (1 - \alpha)$. Suppose $-L > U$. Then $I_{1 - \gamma} (\delta^2)$ also includes values of $\delta$ for which $U < \delta < -L$, which were not included in the initial level-$(1 - \alpha)$ CI for $\delta$. Thie extra coverage  probability is
 \begin{equation} \label{eq:extraLU2}
 \begin{array}{ll}
 P(U < \delta < -L) & = P(\hat\delta -c_{\alpha-\beta} v(\hat\delta) <\delta < -\hat\delta + c_{1 - \beta}v(\hat\delta))\\
                              & = P( (\hat\delta - \delta) / v(\hat\delta) < \min( c_{\alpha-\beta}, c_{1 - \beta} - 2 \delta / v(\hat\delta))\\
                              & = \min(G( c_{\alpha - \beta}), G(c_{1 - \beta} - 2 \delta / v(\hat\delta))).
 \end{array}
 \end{equation}
When $\beta = \alpha / 2 = .025$, the probability $P(U < \delta < -L)$  decreases from
 .025 when $\delta / v(\hat\delta) < c_{1 - \alpha/2} $ to 1.84e-04 when $\delta / v(\hat\delta) = 1.5 c_{1 - \beta} / 2$ to 1.242e-08 when $\delta / v(\hat\delta) = 2c_{1 - \beta}$.

To obtain exactly  $(1 - \alpha)$-level coverage when   $L < 0 < U$, we can choose a value of $\beta$  $0 < \beta < \alpha$ so that $- L = U$ and take the corrected interval $(0, L^2)$ as a level $(1 - \alpha)$ CI for $\delta^2$.  This is equivalent to finding $\beta$ such that $c_{1 - \beta} + c_{\alpha - \beta} = 2 \hat\delta / v(\hat\delta)$. This equation always has a solution: when $\beta \to \alpha$, $c_{1 - \beta} + c_{\alpha -\beta} \to -\infty$, and when $\beta \to 0$, $c_{1 - \beta} + c_{ \alpha - \beta} \to \infty$.

Our simulations included the  above correction to the na\"ive  confidence interval for $L < 0 < U$.
% This corrected CI  provided coverage above 93.5\% in all configurations (see results in the preprint). However, the na\"ive confidence interval provided, overall,  better coverage.

\subsection{Conditional inference for $\delta^2$ given $\hat\tau^2$}\label{sec:cond}

Section~\ref{sec:delta2_orig} suggests the point estimate $ \widehat{\widehat{\delta^2}} = \widehat{\delta^2} - \hat \tau^2$ for the magnitude effect (conditional on $\hat{\tau}^2$). Obtaining a confidence interval for $\delta^2$ given $\hat\tau^2$ is more complicated because $\widehat\delta^2$ and $\hat \tau^2$ are not independent. A simple way forward uses Equation~(\ref{eq:REMmarg}) and the statistic
\begin{equation} \label{eq:REMnc}
\Lambda(\tau^2) = \sum \frac{\tilde{n}_i d_i^2}{1 + \tilde n_i \tau^2} \sim \sum F_{1,m_i} \left(\frac{\tilde{n}_i \delta^2} {1 + \tilde n_i \tau^2} \right)
\underset{\{m_i\} \to \infty} {\sim} \chi^2_K \left( \sum \frac{\tilde{n}_i \delta^2} {1 + \tilde n_i \tau^2} \right).
\end{equation}
%When $m_i\to \infty$, the distribution in (\ref{eq:REMnc}) converges to $\chi^2_K\big(\sum\frac{\tilde{n}_i \delta^2}{1+\tilde n_i\tau^2}\big)$.
A conditional (given $\hat\tau^2$) test for $\delta^2 = 0$ would compare $\Lambda(\hat\tau^2)$ against a percentile from the $\chi^2_K$ distribution, or a critical value obtained by bootstrapping the distribution of $\sum F_{1,m_i}$.
In the same vein, to obtain a conditional (given $\hat\tau^2$) $\chi^2$-profile confidence interval for $\delta^2$,
we can substitute $\hat\tau^2$ for $\tau^2$ in Equation~(\ref{eq:REMnc}) and solve for the confidence limits for $\delta^2 | \hat\tau^2$ at the .025 and .975 percentage points. %as described at the end of Section 2.

\section{Simulation study} \label{sec:simsect}

\subsection{Simulation design} \label{sec:simdes}

A number of other studies have used simulation to examine estimators of $\tau^2$ or of the overall effect for SMD. Our simulation design largely follows that of \cite{BHK2018SMD}, which includes a detailed summary of previous simulation studies and gives our rationale for choosing the ranges of values for $\mu$, $\delta$, and $\tau^2$ that we consider realistic for a range of applications.

All simulations used the same numbers of studies  ($K = 5, \;10, \;20, \;30, \;50, \;100$) and,  for each combination of parameters, the same vector of total sample sizes $(n_{1},\ldots, n_{K})$ and the same proportion of observations in the Control arm ($f_i = .5$ for all $i$). Thus, the sample sizes in the Treatment and Control arms were approximately equal: $n_{iT} = \lceil{n_{i} / 2} \rceil$ and $n_{iC} = n_{i} - n_{iT}$, $i = 1,\ldots, K$.
%The values of $q$ reflect two situations for the two arms of each study: approximately equal (1:1) and quite unbalanced (1:3).

We studied equal and unequal study sizes. %For equal study sizes $n_i$ was as small as 40, and for unequal study sizes $n_i$ was as small as 12 (???), in order to examine how the methods perform for the extremely small sample sizes that arise in some areas of application.
For equal-sized studies,  the  sample sizes were $n_{i} = 40,\; 100,\; 250,\; 500$. %,\; 1000$.
In choosing unequal study sizes, we followed a suggestion of \cite{sanches-2000}, who selected sets of study sizes having skewness 1.464, which they considered typical in behavioral and health sciences. Table \ref{tab:altdataSMD} gives the details.
%The average study sizes in these simulations are $\bar{n} = 30, \;60, \;100, \;160$.  The average size $\bar{n} = 30$ corresponds to studies of sizes $12, \;16, \;18, \;20, \;84$; $\bar{n} = 60$ corresponds to studies of sizes $24, \;32, \;36, \;40, \;168$; $\bar{n} = 100$ corresponds to $64, \;72, \;76, \;80, \;208$; and $\bar{n} = 160$ corresponds to $124, \;132, \;136, \;140, \;268$. For  meta-analyses with $K=10$ and $K=30$,  the same set of sample sizes is used twice or six times, respectively.

We used a total of $10,000$ repetitions for each combination of parameters. Thus, the simulation standard error for estimated coverage of $\tau^2$, $\delta$ or $\delta^2$ at the $95\%$ confidence level is roughly $\sqrt{.95 \times .05 / 10,000} = .00218$.

The simulations were programmed in R version 4.0.2.

\begin{table}[ht]
	\caption{\label{tab:altdataSMD} \emph{Data patterns in the simulations for squared SMD}}
	\begin{footnotesize}
		\begin{center}
			\begin{tabular}
				{|l|l|l|}
				\hline

				Squared SMD&Equal study sizes& Unequal study sizes\\
                &&\\	\hline
				$K$ (number of studies)& 5, 10, 20, 30, 50, 100&5, 10, 30\\
				$n$ or $\bar{n}$  (average (individual) study size&40, 100, 250, 500&60 (24,32,36,40,168), \\
				--- total of the two arms)&&100 (64,72,76,80,208),\\
				For  $K=10$ and $K=30$,  the same set of unequal &&160 (124,132,136,140,268) \\
				study sizes was used twice or six times, respectively.&& \\
				$f$ (proportion of observations in the Control arm) & 1/2&1/2\\
				$\delta$ (true value of the SMD) &0, 0.2, 0.5, 1, 2&0, 0.2, 0.5, 1, 2\\
				$\tau^{2}$ (variance of random effects)&0(0.1)1&0(0.1)1\\
				\hline
			%	* results in Web Appendicex C1, C2&&\\
			%	** results in Web Appendicex D1, D2&&\\
			%	\hline
			\end{tabular}
		\end{center}
	\end{footnotesize}
\end{table}

We varied four parameters: the overall true SMD ($\delta$), the between-studies variance ($\tau^2$), the number of studies ($K$), and the studies' total sample size ($n$ and $\bar{n}$). Table~\ref{tab:altdataSMD} lists the values of each parameter.

We generated the true effect sizes $\delta_{i}$ from a normal distribution:  $\delta_{i} \sim N(\delta, \tau^2)$. We generated the values of  ${d}_{i}$ directly from the appropriately scaled noncentral $t$-distribution, $\tilde n_i^{1/2} d_i \sim t_{m_i} (\tilde{n}_i^{1/2} \delta_i)$, and obtained the values of Hedges's $g_i$ and $d_i^2$ for further meta-analysis of SMD and of ASMD, respectively.

\subsection{Tests and estimators studied}\label{sec:simest}

Under the random-effects model for SMD, we used the generated values of Hedges's $g_i$ to calculate the three  estimators of $\tau^2$ (MP, KDB, and SSC) that \cite{BHK2018SMD} and \cite{BHK2022QSMD} recommended as the best available.  Briefly, \cite{mandel1970interlaboratory} (MP) estimator $\hat{\tau}_{MP}^2$ is  based on the first moment of the large-sample chi-square distribution $\chi_{K-1}^2$ of Cochran's $Q$. \cite{kulinskaya2011testing} derived  $O(1/n)$ corrections to moments of $Q$. The KDB  estimator $\hat{\tau}_{KDB}^2$ is a moment-based estimator based on this improved approximation.   A generalised $Q$ statistic discussed in \cite{dersimonian2007random} and further studied for SMD by \cite{BHK2018SMD} and \cite{BHK2022QSMD}, allows the weights $w_i$ to be arbitrary positive constants. The SSC estimator $\hat{\tau}_{SSC}^2$ is a moment-based estimator with effective sample size weights $\tilde n$.

As a baseline, we  recorded the  bias of these three estimators and the  bias of the three point estimators of $\delta$ that used the MP, KDB, or SSC estimate of $\tau^2$ in the weights.

Point estimators for $\delta$  are weighted averages of estimated SMDs $g_i$. Estimators  corresponding to MP and KDB ($\hat\delta_{MP}$ and $\hat\delta_{KDB}$ ) use inverse-variance-weights obtained by substitution of MP or KDB estimator of  $\tau^2$ into expression for  an inverse-variance-weights  $w_i(\tau^2)=(v_i^2+\tau^2)^{-1}$.  The SSC point estimator of $\delta$ uses effective sample size weights $\tilde{n}$.

Under the common-effect model for ASMD, we studied bias of $d^2$, empirical levels and power of a chi-square test for $\delta^2 = 0$ based on $(\sum \tilde n_i) d^2$ (Equation~(\ref{eq:CDF_FEM})), and coverage of the chi-square profile confidence interval for $\delta^2$ at the 95\% nominal level.

In random-effects  meta-analysis of ASMD, we studied the bias of three point estimators of $\delta^2$ ($\widehat{\widehat{\delta_{MP}^2}}$, $\widehat{\widehat{\delta_{KDB}^2}}$, and $\widehat{\widehat{\delta_{SSC}^2}}$) calculated as  $\widehat{\widehat{\delta^2_{\hat\tau^2}}} = \widehat{\delta^2} - \hat\tau^2$, where $\widehat{\delta^2}$ is given by   Equation~(\ref{eq:Delta_FEM}) and $\hat\tau^2$ is given by the corresponding estimator of $\tau^2$, and of their truncated-at-zero versions, calculated as  $\widehat{\widehat{\delta^2_{tr}}}=\max(\widehat{\widehat{\delta^2}},0)$.

We studied coverage of the 95\% confidence intervals for $\delta^2$ based on the confidence intervals for the signed $\delta$ values, described in Section \ref{sec:delta2_orig}.
 We considered both na\"ive and corrected versions of these CIs . We used percentage points from the normal distribution for the MP, KDB, and SSC-based intervals and $t_{K-1}$ percentage points for a second SSC-based interval, denoted by SSC\_t.

 Interval estimators for $\delta$ corresponding to MP, KDB and SSC use the respective point estimator $\hat\delta$ as the midpoint, and the half-width equals the estimated standard deviation of $\hat\delta$  under the random-effects model times the critical value from the normal or (for SSC\_t) from the $t$ distribution on $K - 1$ degrees of freedom.

We also studied coverage of the three conditional 95\% confidence intervals for $\delta^2$ $\Lambda_{MP}$, $\Lambda_{KDB}$, and $\Lambda_{SSC}$ based on the statistic $\Lambda(\tau^2)$ given by Equation~(\ref{eq:REMnc}) in combination with the estimates  $\tau_{MP}^2$, $\tau_{KDB}^2$, and $\tau_{SSC}^2$.

Additionally, we  studied empirical levels and power of the conditional tests of $\delta^2=0$ based on the statistics $\Lambda_{MP}$, $\Lambda_{KDB}$, and $\Lambda_{SSC}$ and the $\sum F_{1,m_i}$ distribution or the $\chi^2_K$ approximation to this  distribution (Equation~(\ref{eq:REMnc})). For comparison, we also studied empirical levels of the unconditional test based on $\Lambda(\tau^2)$ for known $\tau^2$.

% Finally, we also attempted to study bias and coverage in MLE  estimation of $|\delta|$ and $\tau^2$ using the folded normal approximation to the distribution %of $|d_i|$.

\section{Simulation results}
\subsection{Baseline estimation of $\delta$ and $\tau^2$}

In estimation of $\delta$, the maximum average bias across all configurations  was below $0.01$, and the median bias was $-0.002$ or less for all three estimators. \\%$0.00865,\;	0.00865,\;	0.00971$ for $\hat\theta_{MP}$,$\hat\theta_{SSC}$, respectively. median bias -0.002269707	-0.002257417	-1.12666E-05
 In estimation of $\tau^2$, the maximum bias was higher, at 0.045 or less,  but it decreased to 0.017 or less for $n\geq 100$.  The median bias was less than 0.0015.
%-0.001238509	0.001564433	0.000312735 across all combinations.
\cite{BHK2018SMD,BHK2022QSMD} give more details on the behavior of our chosen estimators.

\subsection{Bias of point estimators of $\delta^2$, Appendix A}

When $\delta=0$, all three estimators had a small negative bias for $\tau^2\leq 0.1$, but were almost unbiased for $\tau^2\geq 0.2$. The truncated versions had positive bias, especially pronounced for $K=5$, that increased with increasing $\tau^2$.  SSC was almost unbiased. For larger values of $\delta$, bias varied more among the estimators when $n=40$ and $K=5$. However, for larger $n$, the bias of all estimators was very small.

\subsection{Empirical levels of the conditional tests of $\delta^2=0$, Appendix B}

All three conditional tests of $\delta^2=0$ at a 5\% nominal level proved unfit for use. The levels were near zero when $\tau^2=0$, but when $K=5$, they increased to near nominal for $\tau^2=0.2$ and increased to about 0.06 by $\tau^2=1$. The tests based on the bootstrap $F$ values behaved similarly, with somewhat lower levels. However, for $K = 10$, the levels increased to about 0.02 and remained there, and for $K = 20$ they were near zero for all $\tau^2$ values. In contrast, the unconditional test, which used known $\tau^2$, produced consistent near-nominal levels. We believe that the disappointing behavior of the conditional tests arises from high correlation between the $d_i^2$ %$\Lambda(\tau^2)$
and $\hat\tau^2$ values. This correlation is well known for the folded normal distribution \citep{tsagris-2014}.
%, and Equation~(\ref{eq:REMmarg_a}) leads us to expect it.

\subsection{Coverage of na\"ive  and corrected confidence intervals for $\delta^2$ based on signed SMD values, Appendix C}

Coverage did not depend much on sample sizes.  Confidence intervals based on normal critical values generally had low coverage for $K<30$, especially for small $K$ and $\delta = 0.2$ or $0.5$, but their coverage improved with $K$.
There was no visible difference among the MP, KD, or SSC confidence intervals.

Na\"ive SSC\_t confidence intervals, based on $t_{K-1}$ critical values, provided consistently good coverage for the vast majority of configurations.  For $\delta = 0$ or $\delta \geq 1$, their coverage was almost nominal for $\tau^2 \geq 0.2$. For $0.5 \geq \delta \geq 0.2$, coverage was above nominal when $K\leq 10$, but for $K \geq 20$ it decreased to nominal  for $\delta = 0.5$.
Even for $K=100$, coverage was somewhat above nominal for large $\tau^2$ values when $\delta = 0.2$.

For $K \geq 50$, there was almost no difference in coverage between normal- and t-based intervals.

We also studied coverage of the corrected confidence intervals. Coverage of the corrected SSC*\_t confidence intervals was above 93.5\% for all configurations, but it was typically below nominal for $\delta = 0.2$ and $0.5$, even for $K = 100$. Therefore, we do not recommend this correction.

\subsection{Coverage of conditional confidence intervals for $\delta^2$, Appendix C}

When $\delta=0$, coverage of the conditional confidence intervals follows from the above results on the empirical levels of the respective conditional tests.
There was not much difference among the MP, KD, and SSC conditional  confidence intervals, nor among sample sizes from $n=40$ to $n=1000$.
For $K = 5$, coverage was near 1 when $\tau^2 = 0$, it slowly decreased to nominal for larger $\tau^2$. For $K = 10$, coverage decreased from 1 to about 98\%, and for $K \geq 20$, coverage was near 1 for all $\tau^2$ values. However, for larger values of $\delta$, coverage was near nominal when $\tau^2 = 0$ and then dropped dramatically for larger $\tau^2$. This drop was more pronounced for $K \leq 10$ and for larger $\delta$. It was quite prominent when $K = 5$ and $\delta = 0.5$ but less so for $K = 30$ and $\delta = 0.5$, where it was above nominal, but the drop was present even when $K = 100$ and $\delta = 1$. Coverage then increased slowly with increasing $\tau^2$, sometimes almost to nominal when $\tau^2 = 1$. When $\delta = 2$, coverage was low for $\tau^2 = 0$ and increased slowly with $\tau^2$.

\section{Discussion}

Though common in ecology and evolutionary biology, meta-analysis of magnitude effects has received little statistical attention, and the methods used so far are not appropriate. We formulate a random-effects model for meta-analysis of ASMD and propose appropriate statistical methods for point and interval estimation in meta-analysis of ASMD.

Statistical properties of squared SMD are more straightforward than those of its absolute value. Therefore, our methodological development focuses mainly on inference for $\delta^2$. However, for inference on $|\delta|$, one only needs to take the square root of the estimated $\delta^2$ and its confidence limits.

For point estimation of squared  ASMD, we corrected an estimate of $\delta^2$ by subtracting the estimated  between-study variance $\hat{\tau}^2$ (in the signed SMD meta-analysis). Our simulations show that this works well when using a good estimator of $\tau^2$ such as MP, KD, or SSC.

For interval estimation, we considered three classes of statistical methods: na\"ive and corrected  intervals for $\delta^2$ obtained from the  signed SMD data and  conditional methods based on the distribution of $\delta^2$ given the estimated  $\tau^2$. %, and an approximation based on the folded normal distribution.
We found that coverage of the conditional confidence intervals was rather erratic, and the corrected confidence intervals provided somewhat low coverage in the vicinity of zero. %, probably because of strong correlations between quadratic forms in estimated %effects and $\tau^2$.  The methods based on the folded normal %distribution performed very badly, in both point and interval estimation.
However, na\"ive
 %straightforward correction of the confidence levels used for
squaring of the SMD confidence limits, obtained  with percentage points from the $t_{K-1}$ distribution, provided reliable coverage across all configurations  of the parameters in our simulations and can be recommended for use in practice.

%\section*{Data Availability Statement}
% The user-friendly R program implementing
%all studied estimators of absolute standardised mean differences $|\delta|$ in meta-analysis of magnitude effects  with related confidence intervals is %available  at https://osf.io/5n3vd.

\section*{Acknowledgements}

We are grateful to Prof Julia Koricheva who brought the meta-analysis of magnitude effects to our attention.\\
We would also like to thank Dr Michael Tsagris who kindly provided his simulation program for MLE estimation of parameters of the folded normal distribution used in \cite{tsagris-2014} and recommended the use of \textit{Rfast} R package for this purpose. \\
The work by E. Kulinskaya was supported by the Economic and Social Research Council
[grant number ES/L011859/1].

\bibliography{asmd1.bib}

\begin{thebibliography}{24}
\providecommand{\natexlab}[1]{#1}
\providecommand{\url}[1]{\texttt{#1}}
\expandafter\ifx\csname urlstyle\endcsname\relax
  \providecommand{\doi}[1]{doi: #1}\else
  \providecommand{\doi}{doi: \begingroup \urlstyle{rm}\Url}\fi

\bibitem[Ali et~al.(2019)Ali, Prieto-Alhambra, Lopes, Ramos, Bispo, Ichihara,
  Pescarini, Williamson, Fiaccone, Barreto, and Smeeth]{Sanni-2019}
M~Sanni Ali, Daniel Prieto-Alhambra, Luciane~Cruz Lopes, Dandara Ramos, Nivea
  Bispo, Maria~Y. Ichihara, Julia~M. Pescarini, Elizabeth Williamson,
  Rosemeire~L. Fiaccone, Mauricio~L. Barreto, and Liam Smeeth.
\newblock Propensity score methods in health technology assessment: Principles,
  extended applications, and recent advances.
\newblock \emph{Frontiers in Pharmacology}, 10\penalty0 (973), 2019.
\newblock \doi{10.3389/fphar.2019.00973}.

\bibitem[Bailey et~al.(2009)Bailey, Schweitzer, Úbeda, Koricheva, LeRoy,
  Madritch, Rehill, Bangert, Fischer, Allan, and Whitham]{Bailey2009}
Joseph~K. Bailey, Jennifer~A. Schweitzer, Francisco Úbeda, Julia Koricheva,
  Carri~J. LeRoy, Michael~D. Madritch, Brian~J. Rehill, Randy~K. Bangert,
  Dylan~G. Fischer, Gerard~J. Allan, and Thomas~G. Whitham.
\newblock From genes to ecosystems: a synthesis of the effects of plant genetic
  factors across levels of organization.
\newblock \emph{Philosophical Transactions of the Royal Society B: Biological
  Sciences}, 364\penalty0 (1523):\penalty0 1607--1616, 2009.
\newblock \doi{10.1098/rstb.2008.0336}.

\bibitem[Bakbergenuly et~al.(2020)Bakbergenuly, Hoaglin, and
  Kulinskaya]{BHK2018SMD}
Ilyas Bakbergenuly, David~C. Hoaglin, and Elena Kulinskaya.
\newblock Estimation in meta-analyses of mean difference and standardized mean
  difference.
\newblock \emph{Statistics in Medicine}, 39\penalty0 (2):\penalty0 171--191,
  2020.
\newblock \doi{10.1002/sim.8422}.

\bibitem[Bakbergenuly et~al.(2022)Bakbergenuly, Hoaglin, and
  Kulinskaya]{BHK2022QSMD}
Ilyas Bakbergenuly, David~C. Hoaglin, and Elena Kulinskaya.
\newblock On the {$Q$} statistic with constant weights for standardized mean
  difference.
\newblock \emph{British Journal of Mathematical and Statistical Psychology},
  75\penalty0 (3):\penalty0 444--465, 2022.
\newblock \doi{10.1111/bmsp.12263}.

\bibitem[Champagne et~al.(2016)Champagne, Tremblay, and Côté]{Champagne2016}
Emilie Champagne, Jean-Pierre Tremblay, and Steeve~D. Côté.
\newblock Spatial extent of neighboring plants influences the strength of
  associational effects on mammal herbivory.
\newblock \emph{Ecosphere}, 7\penalty0 (6):\penalty0 e01371, 2016.
\newblock \doi{10.1002/ecs2.1371}.

\bibitem[Clements et~al.(2022)Clements, Sundin, Clark, and
  Jutfelt]{Clements2022}
Jeff~C. Clements, Josefin Sundin, Timothy~D. Clark, and Fredrik Jutfelt.
\newblock Meta-analysis reveals an extreme “decline effect” in the impacts
  of ocean acidification on fish behavior.
\newblock \emph{PLoS Biology}, 20\penalty0 (2):\penalty0 e3001511, 02 2022.
\newblock \doi{10.1371/journal.pbio.3001511}.

\bibitem[Costantini(2018)]{Costantini2017}
David Costantini.
\newblock Meta-analysis reveals that reproductive strategies are associated
  with sexual differences in oxidative balance across vertebrates.
\newblock \emph{Current Zoology}, 64\penalty0 (1):\penalty0 1--11, 2018.
\newblock \doi{10.1093/cz/zox002}.

\bibitem[DerSimonian and Kacker(2007)]{dersimonian2007random}
Rebecca DerSimonian and Raghu Kacker.
\newblock Random-effects model for meta-analysis of clinical trials: an update.
\newblock \emph{Contemporary Clinical Trials}, 28\penalty0 (2):\penalty0
  105--114, 2007.
\newblock \doi{10.1016/j.cct.2006.04.004}.

\bibitem[Felix et~al.(2023)Felix, Stevenson, and Koricheva]{Koricheva2023}
Juri~A. Felix, Philip~C. Stevenson, and Julia Koricheva.
\newblock Plant neighbourhood diversity effects on leaf traits: A
  meta-analysis.
\newblock \emph{Functional Ecology}, in press, 2023.
\newblock \doi{10.1111/1365-2435......}

\bibitem[Garamszegi(2006)]{Garamszegi2006}
László~Zsolt Garamszegi.
\newblock {Comparing effect sizes across variables: generalization without the
  need for Bonferroni correction}.
\newblock \emph{Behavioral Ecology}, 17\penalty0 (4):\penalty0 682--687, 2006.
\newblock \doi{10.1093/beheco/ark005}.

\bibitem[Gelman(2006)]{Gelman-2006}
Andrew Gelman.
\newblock Prior distributions for variance parameters in hierarchical models
  (comment on article by {B}rowne and {D}raper).
\newblock \emph{Bayesian Analysis}, 1\penalty0 (3):\penalty0 515--534, 2006.
\newblock \doi{10.1214/06-BA117A}.

\bibitem[Hedges(1983)]{hedges1983random}
Larry~V. Hedges.
\newblock A random effects model for effect sizes.
\newblock \emph{Psychological Bulletin}, 93\penalty0 (2):\penalty0 388--395,
  1983.

\bibitem[Johnson et~al.(1995)Johnson, Kotz, and Balakrishnan]{J-K-B-1995}
Norman~L. Johnson, Samuel Kotz, and N.~Balakrishnan.
\newblock \emph{Continuous {U}nivariate {D}istributions, {V}olume 2}.
\newblock John Wiley \& Sons, New York, second edition, 1995.

\bibitem[Jones and Marchand(2021)]{jones2021}
M.~C. Jones and Éric Marchand.
\newblock A (non-central) chi-squared mixture of non-central chi-squareds is
  (non-central) chi-squared and related results, corollaries and applications.
\newblock \emph{Stat}, 10\penalty0 (1):\penalty0 e398, 2021.
\newblock \doi{10.1002/sta4.398}.

\bibitem[Kulinskaya et~al.(2004)Kulinskaya, Dollinger, Knight, and
  Gao]{kulinskaya2004welch}
E.~Kulinskaya, M.~B. Dollinger, E.~Knight, and H.~Gao.
\newblock A {W}elch-type test for homogeneity of contrasts under
  heteroscedasticity with application to meta-analysis.
\newblock \emph{Statistics in Medicine}, 23\penalty0 (23):\penalty0 3655--3670,
  2004.
\newblock \doi{10.1002/sim.1929}.

\bibitem[Kulinskaya et~al.(2011)Kulinskaya, Dollinger, and
  Bj{\o}rkest{\o}l]{kulinskaya2011testing}
Elena Kulinskaya, Michael~B. Dollinger, and Kirsten Bj{\o}rkest{\o}l.
\newblock Testing for homogeneity in meta-analysis {I}. {T}he one-parameter
  case: standardized mean difference.
\newblock \emph{Biometrics}, 67\penalty0 (1):\penalty0 203--212, 2011.
\newblock \doi{10.1111/j.1541-0420.2010.01442.x}.

\bibitem[Leone et~al.(1961)Leone, Nelson, and Nottingham]{Leone1961}
F.~C. Leone, L.~S. Nelson, and R.~B. Nottingham.
\newblock The folded normal distribution.
\newblock \emph{Technometrics}, 3\penalty0 (4):\penalty0 543--550, 1961.
\newblock \doi{10.1080/00401706.1961.10489974}.

\bibitem[Mandel and Paule(1970)]{mandel1970interlaboratory}
John Mandel and Robert~C. Paule.
\newblock Interlaboratory evaluation of a material with unequal numbers of
  replicates.
\newblock \emph{Analytical Chemistry}, 42\penalty0 (11):\penalty0 1194--1197,
  1970.

\bibitem[Morrissey(2016)]{morrisey-2016}
M.~B. Morrissey.
\newblock Meta-analysis of magnitudes, differences and variation in
  evolutionary parameters.
\newblock \emph{Journal of Evolutionary Biology}, 29\penalty0 (10):\penalty0
  1882--1904, 2016.
\newblock \doi{10.1111/jeb.12950}.

\bibitem[Psarakis and Panaretoes(1990)]{psarakis-1990}
S.~Psarakis and J.~Panaretoes.
\newblock The folded t distribution.
\newblock \emph{Communication in Statistics--Theory and Methods}, 19\penalty0
  (7):\penalty0 2717--2734, 1990.
\newblock \doi{10.1080/03610929008830342}.

\bibitem[Rubin(2001)]{rubin-2001}
D.~B. Rubin.
\newblock Using propensity scores to help design observational studies:
  {A}pplication to the tobacco litigation.
\newblock \emph{Health Services \& Outcomes Research Methodology}, 2:\penalty0
  169--188, 2001.

\bibitem[S{\'a}nchez-Meca and Mar{\'\i}n-Mart{\'\i}nez(2000)]{sanches-2000}
Julio S{\'a}nchez-Meca and Fulgencio Mar{\'\i}n-Mart{\'\i}nez.
\newblock Testing the significance of a common risk difference in
  meta-analysis.
\newblock \emph{Computational Statistics \& Data Analysis}, 33\penalty0
  (3):\penalty0 299--313, 2000.

\bibitem[Steiger(2004)]{steiger2004}
James~H. Steiger.
\newblock Beyond the {F} test: Effect size confidence intervals and tests of
  close fit in the analysis of variance and contrast analysis.
\newblock \emph{Psychological Methods}, 9\penalty0 (2):\penalty0 164--182,
  2004.
\newblock \doi{10.1037/1082-989X.9.2.164}.

\bibitem[Tsagris et~al.(2014)Tsagris, Beneki, and Hassani]{tsagris-2014}
M.~Tsagris, C.~Beneki, and H.~Hassani.
\newblock On the folded normal distribution.
\newblock \emph{Mathematics}, 2\penalty0 (10):\penalty0 12--28, 2014.
\newblock \doi{10.3390/math2010012}.

\end{thebibliography}

\section*{Appendices}
\begin{itemize}
\item Appendix A: Bias in point estimation of $\delta^2$
\item Appendix B: Empirical level of conditional tests of $\delta^2=0$ at 5\% nominal level
\item Appendix C: Coverage of 95\% confidence intervals for $\delta^2$
\end{itemize}

\setcounter{figure}{0}
\setcounter{section}{0}
\clearpage

\section*{Appendix A: Bias in point estimation of $\delta^2$}

Each figure corresponds to a value of the standardized mean difference $\delta$ (=0, 0.2, 0.5, 1, 2) and one of three combinations of $n$ or $\bar{n}$ and a set of values of $K$: $n \in$ \{40, 100, 250, 500\} and $K \in$ \{5, 10, 20\},  $n \in$ \{40, 100, 250, 500\} and $K \in$ \{30, 50, 100\}, $\bar{n} \in$ \{60, 100, 160\} and $K \in$ \{5, 10, 20\}.
The fraction of each study's sample size in the Control arm ($f$) is held constant at 0.5.

For each combination of a value of $n$ (=  40, 100, 250, 500) or  $\bar{n}$ (= 60, 100, 160) and a value of $K$ (= 5, 10, 20 or 30, 50, 100), a panel plots bias versus $\tau^2$ (= 0(0.1)1).\\
The point estimators of $\delta^2$ are
\begin{itemize}
\item KDB (Kulinskaya-Dollinger-Bj{\o}rkest{\o}l) method, inverse-variance weights
\item MP (Mandel-Paule) method, inverse-variance weights
\item SSC method, effective-sample-size weights

For each method we include both truncated and non-truncated versions.
\end{itemize}

\clearpage

\setcounter{figure}{0}
\setcounter{section}{0}
\renewcommand{\thefigure}{A.\arabic{figure}}

\begin{figure}[ht]
	\centering
	\includegraphics[scale=0.33]{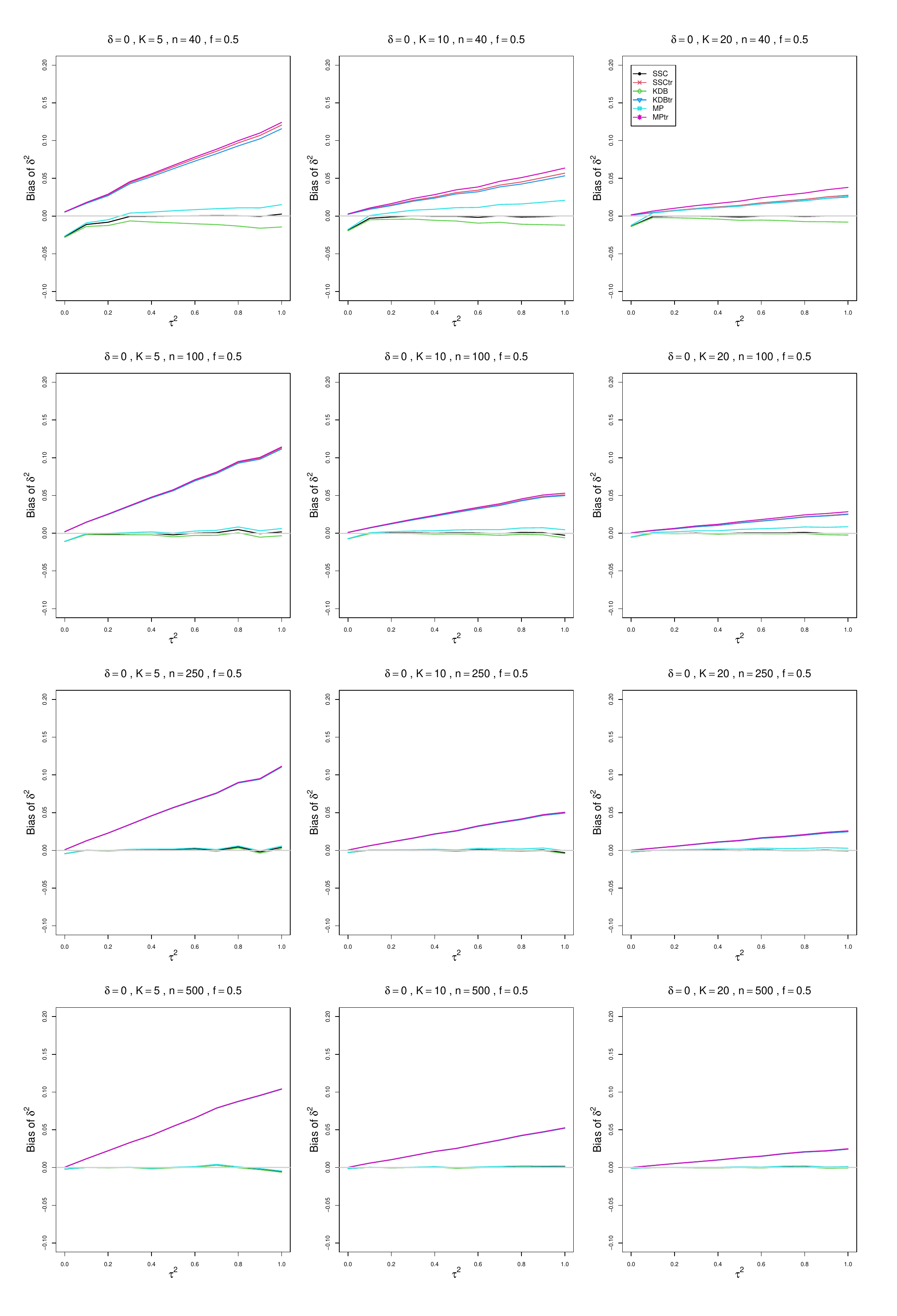}
	\caption{Bias of point estimators of $\delta^2$ based on  MP, KD  and  SMC estimators of $\tau^2$ and their truncated versions vs $\tau^2$, for equal sample sizes $n = 40, \;100,\;250,\;500$, $K=5,\;10$ and $20$, $\delta = 0$ and  $f = 0.5$. }
	\label{PlotBiasOfDelta0Ksmall_equal_sample_sizes.pdf}
\end{figure}

\begin{figure}[ht]
	\centering
	\includegraphics[scale=0.33]{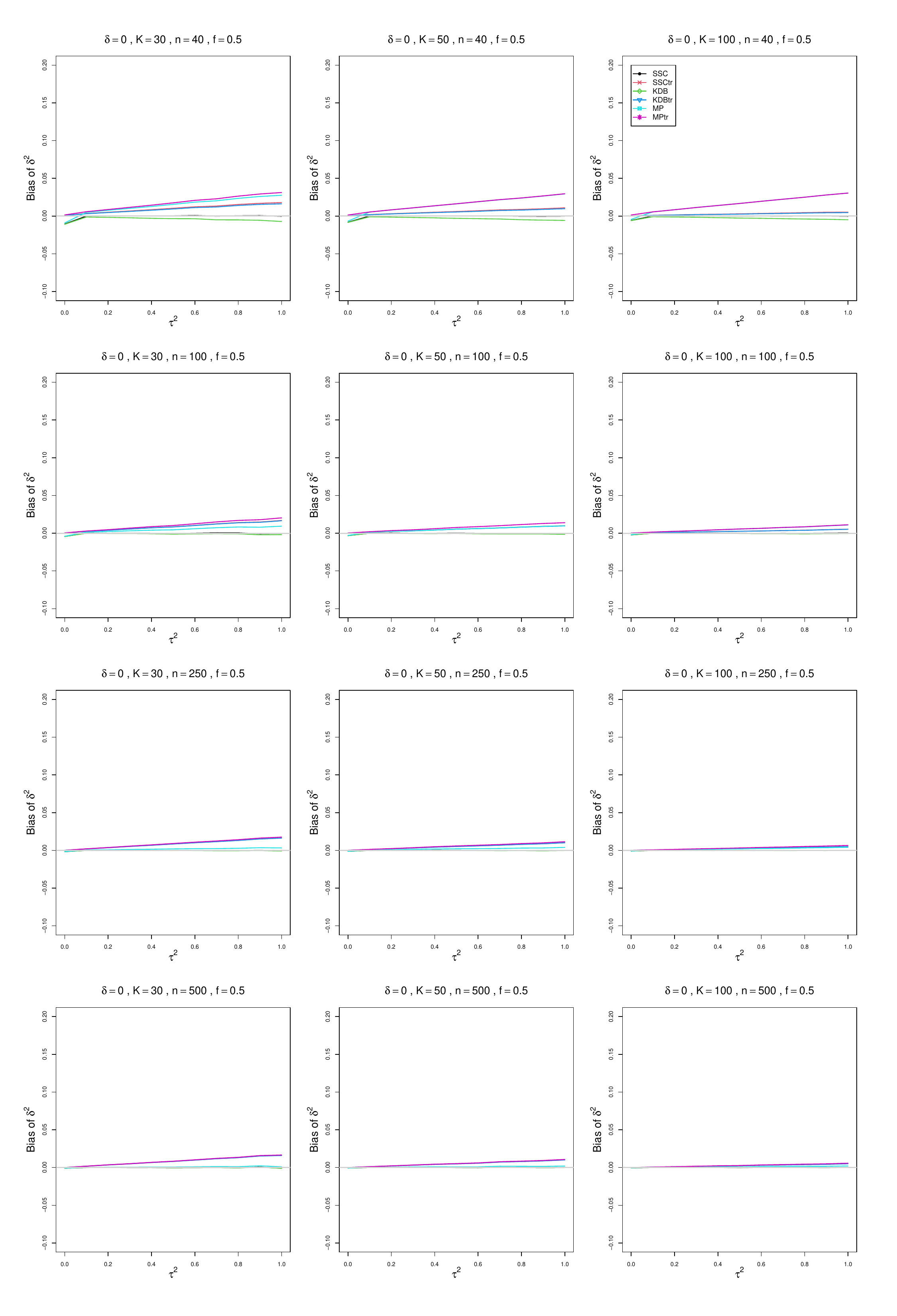}
	\caption{Bias of point estimators of $\delta^2$ based on  MP, KD  and  SMC estimators of $\tau^2$ and their truncated versions vs $\tau^2$, for equal sample sizes $n = 40, \;100,\;250,\;500$, $K=30,\;50$ and $100$, $\delta = 0$ and  $f = 0.5$. }
	\label{PlotBiasOfDelta0Kbig_equal_sample_sizes.pdf}
\end{figure}

\begin{figure}[ht]
	\centering
	\includegraphics[scale=0.33]{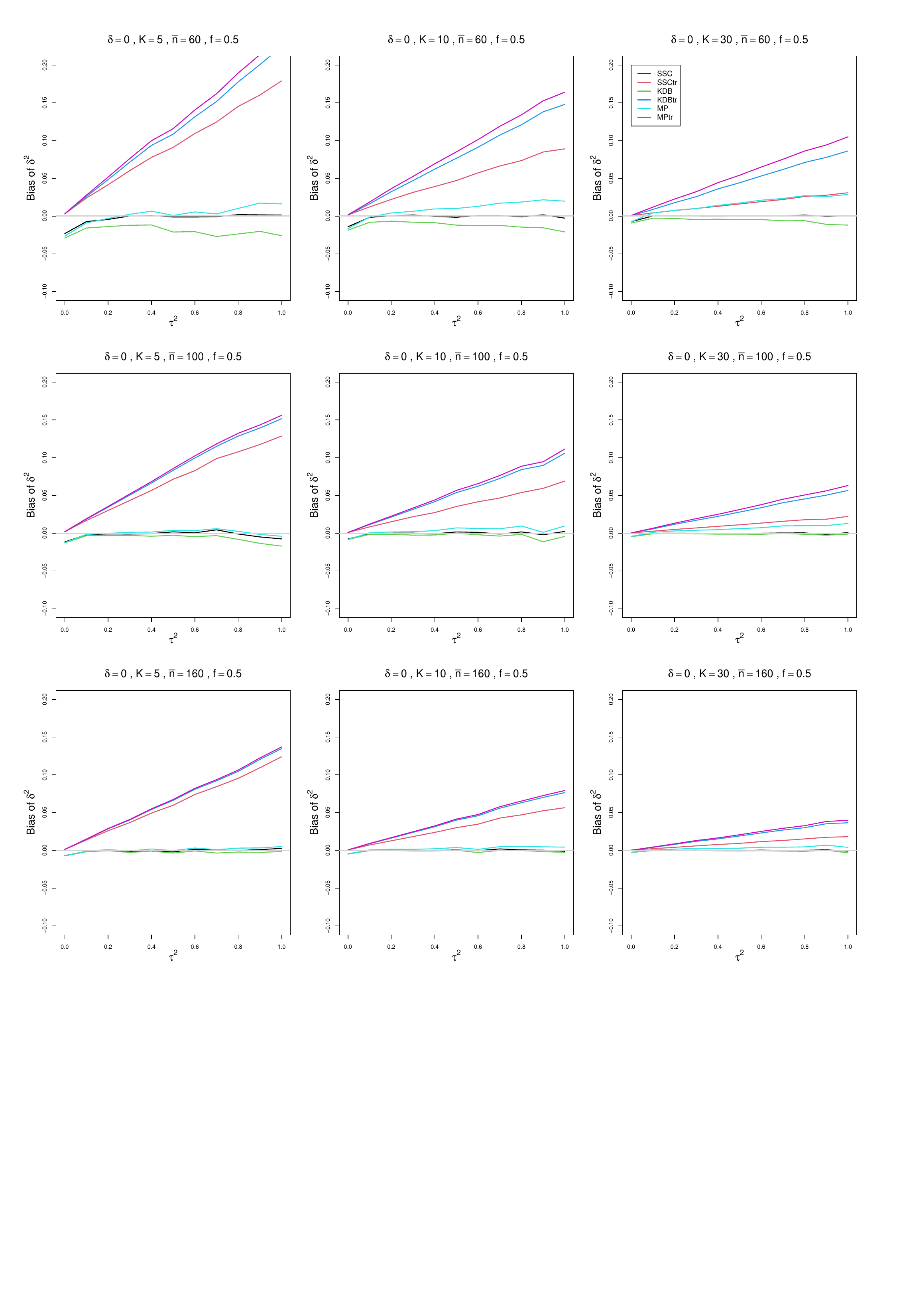}
	\caption{Bias of point estimators of $\delta^2$ based on  MP, KD  and  SMC estimators of $\tau^2$ and their truncated versions vs $\tau^2$, for unequal sample sizes $\bar{n} = 60, \;100,\;160$, $K=5,\;10$ and $30$, $\delta = 0$ and  $f = 0.5$. }
	\label{PlotBiasOfDelta0Kbig_unequal_sample_sizes.pdf}
\end{figure}

\begin{figure}[ht]
	\centering
	\includegraphics[scale=0.33]{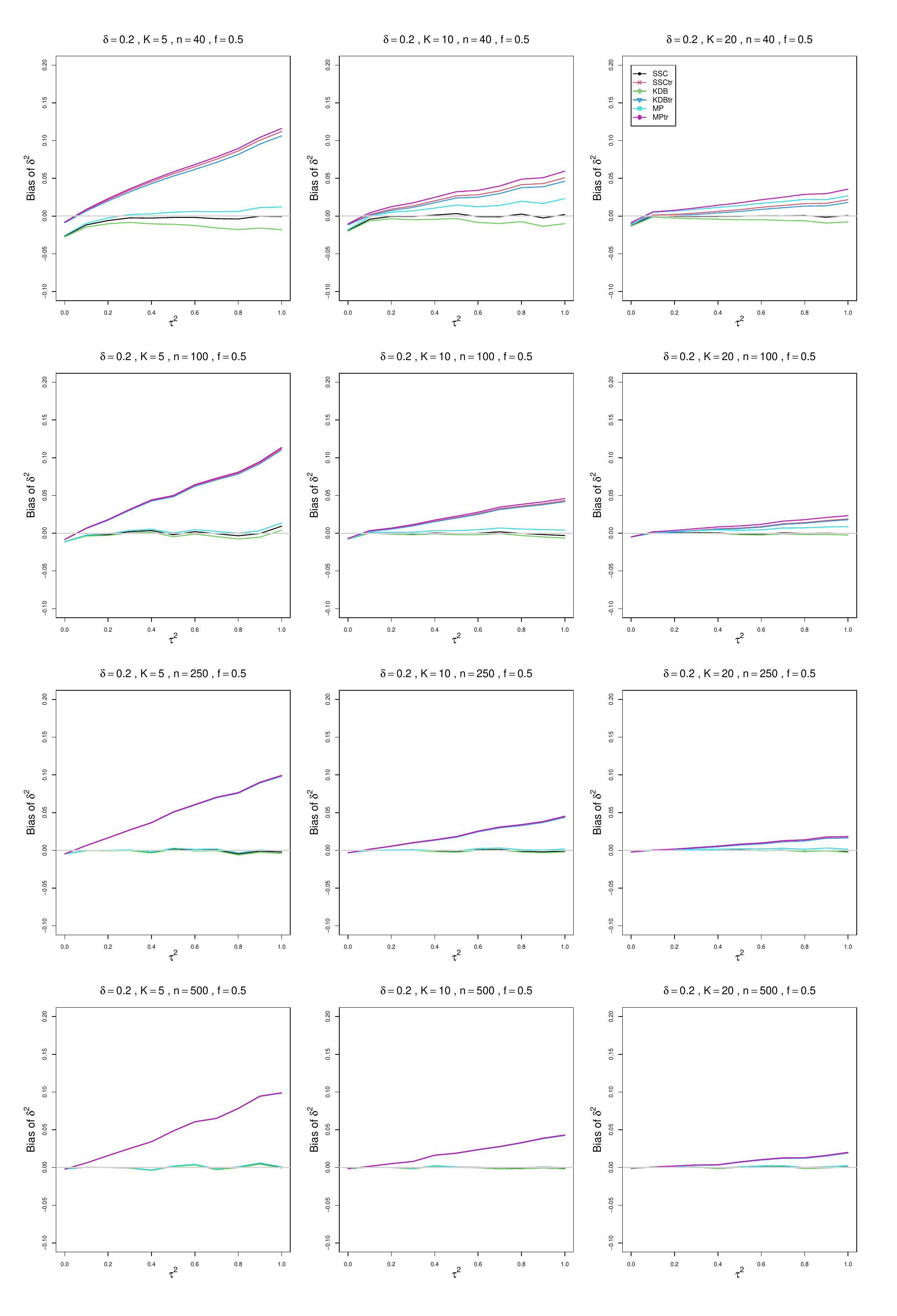}
	\caption{Bias of point estimators of $\delta^2$ based on  MP, KD  and  SMC estimators of $\tau^2$ and their truncated versions vs $\tau^2$, for equal sample sizes $n = 40, \;100,\;250,\;500$, $K=5,\;10$ and $20$, $\delta = 0.2$ and  $f = 0.5$. }
	\label{PlotBiasOfDelta02Ksmall_equal_sample_sizes.pdf}
\end{figure}

\begin{figure}[ht]
	\centering
	\includegraphics[scale=0.33]{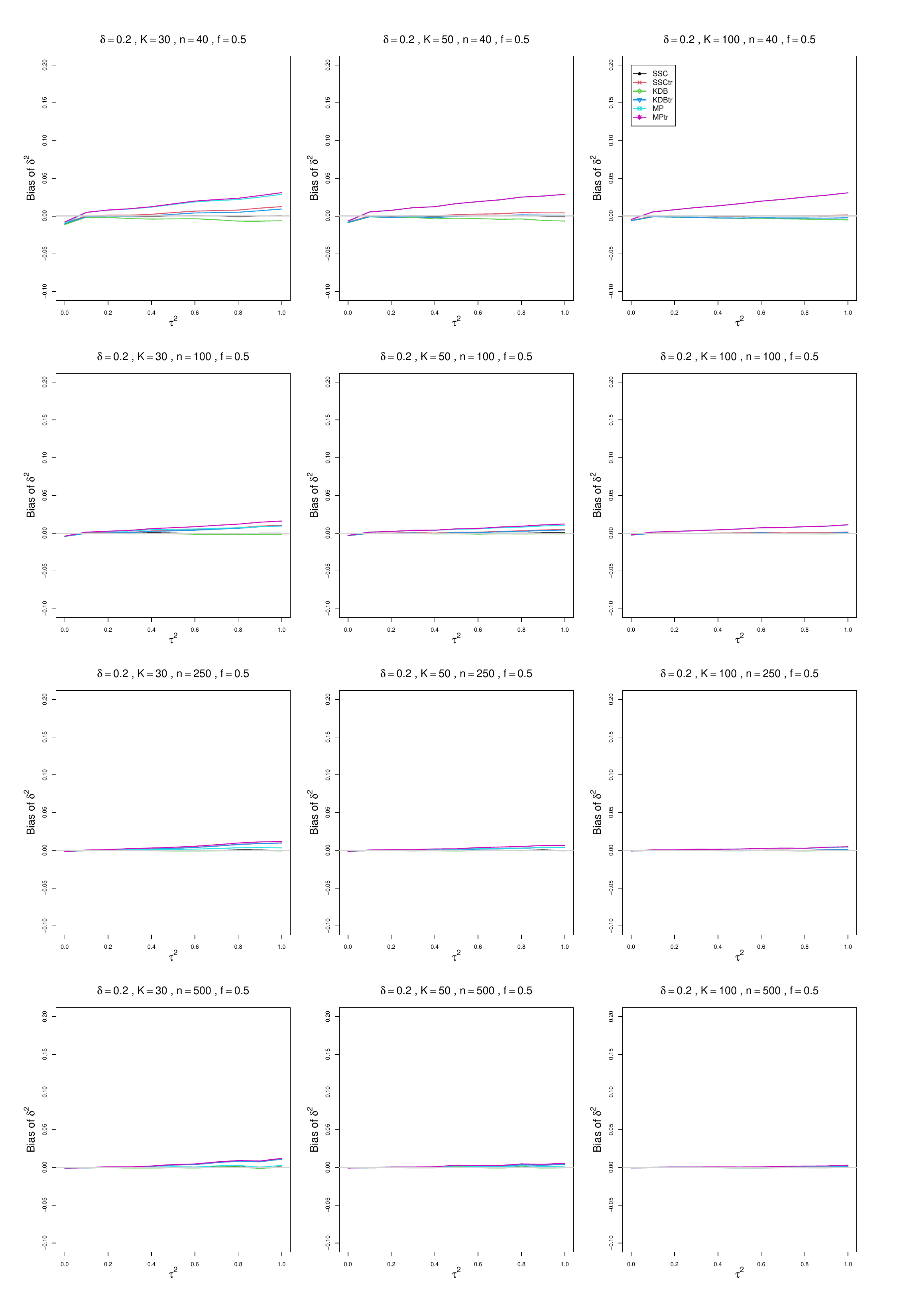}
	\caption{Bias of point estimators of $\delta^2$ based on  MP, KD  and  SMC estimators of $\tau^2$ and their truncated versions vs $\tau^2$, for equal sample sizes $n = 40, \;100,\;250,\;500$, $K=30,\;50$ and $100$, $\delta = 0.2$ and  $f = 0.5$. }
	\label{PlotBiasOfDelta02Kbig_equal_sample_sizes.pdf}
\end{figure}

\begin{figure}[ht]
	\centering
	\includegraphics[scale=0.33]{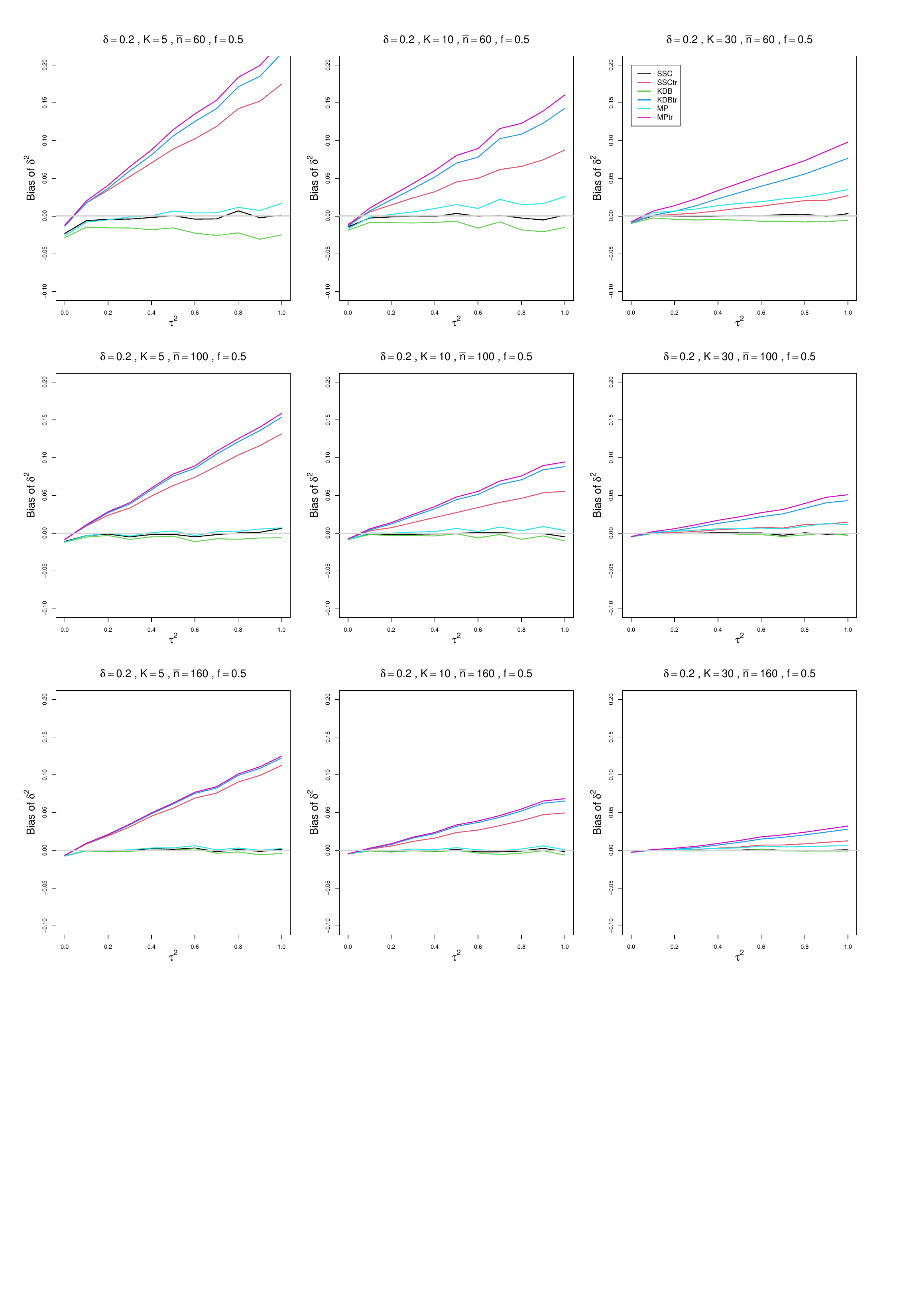}
	\caption{Bias of point estimators of $\delta^2$ based on  MP, KD  and  SMC estimators of $\tau^2$ and their truncated versions vs $\tau^2$, for unequal sample sizes $\bar{n} = 60, \;100,\;160$, $K=5,\;10$ and $30$, $\delta = 0.2$ and  $f = 0.5$. }
	\label{PlotBiasOfDelta02Kbig_unequal_sample_sizes.pdf}
\end{figure}

\begin{figure}[ht]
	\centering
	\includegraphics[scale=0.33]{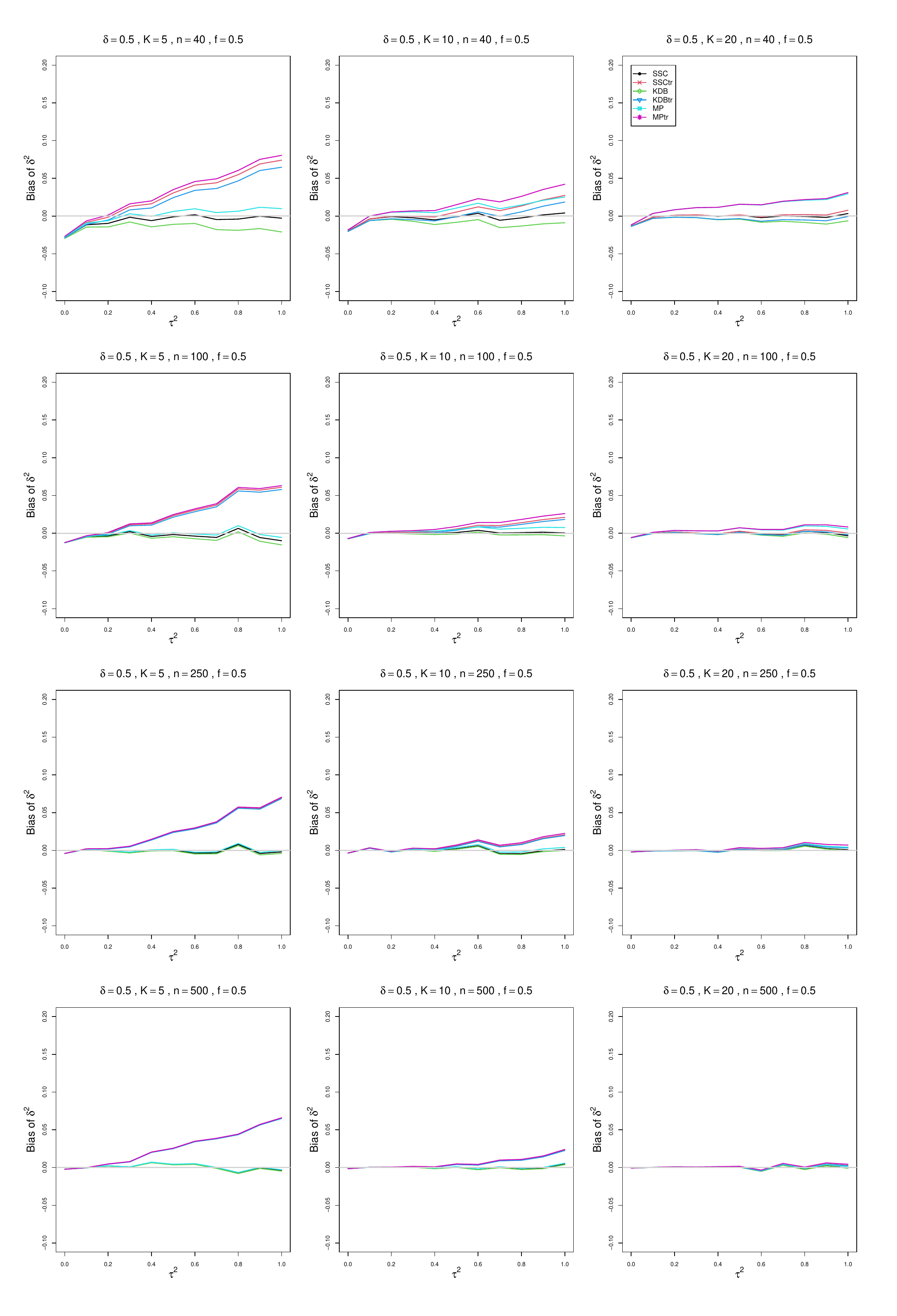}
	\caption{Bias of point estimators of $\delta^2$ based on  MP, KD  and  SMC estimators of $\tau^2$ and their truncated versions vs $\tau^2$, for equal sample sizes $n = 40, \;100,\;250,\;500$, $K=5,\;10$ and $20$, $\delta = 0.5$ and  $f = 0.5$. }
	\label{PlotBiasOfDelta05Ksmall_equal_sample_sizes.pdf}
\end{figure}

\begin{figure}[ht]
	\centering
	\includegraphics[scale=0.33]{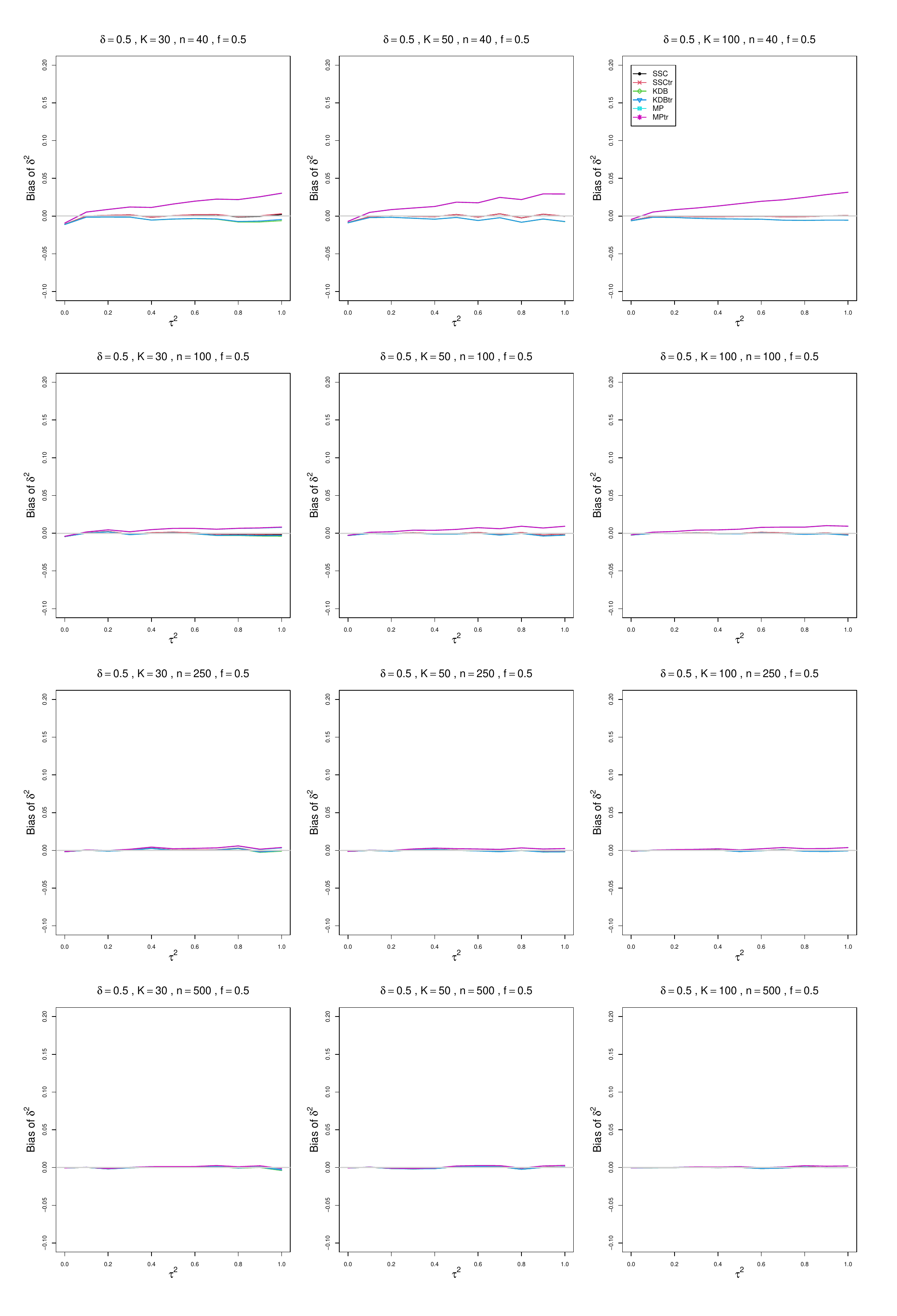}
	\caption{Bias of point estimators of $\delta^2$ based on  MP, KD  and  SMC estimators of $\tau^2$ and their truncated versions vs $\tau^2$, for equal sample sizes $n = 40, \;100,\;250,\;500$, $K=30,\;50$ and $100$, $\delta = 0.5$ and  $f = 0.5$. }
	\label{PlotBiasOfDelta05Kbig_equal_sample_sizes.pdf}
\end{figure}

\begin{figure}[ht]
	\centering
	\includegraphics[scale=0.33]{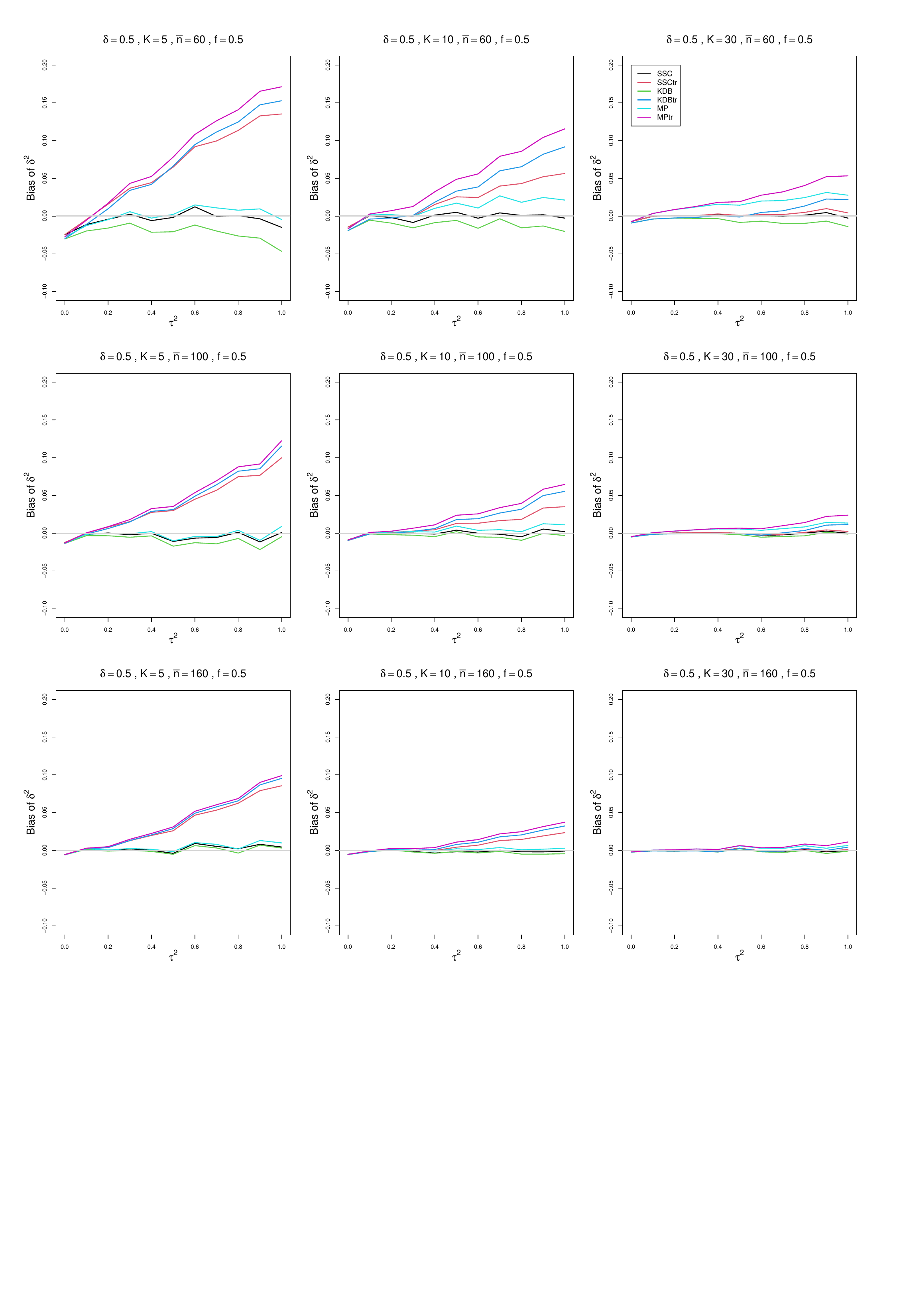}
	\caption{Bias of point estimators of $\delta^2$ based on  MP, KD  and  SMC estimators of $\tau^2$ and their truncated versions vs $\tau^2$, for unequal sample sizes $\bar{n} = 60, \;100,\;160$, $K=5,\;10$ and $30$, $\delta = 0.5$ and  $f = 0.5$. }
	\label{PlotBiasOfDelta05Kbig_unequal_sample_sizes.pdf}
\end{figure}
\begin{figure}[ht]
	\centering
	\includegraphics[scale=0.33]{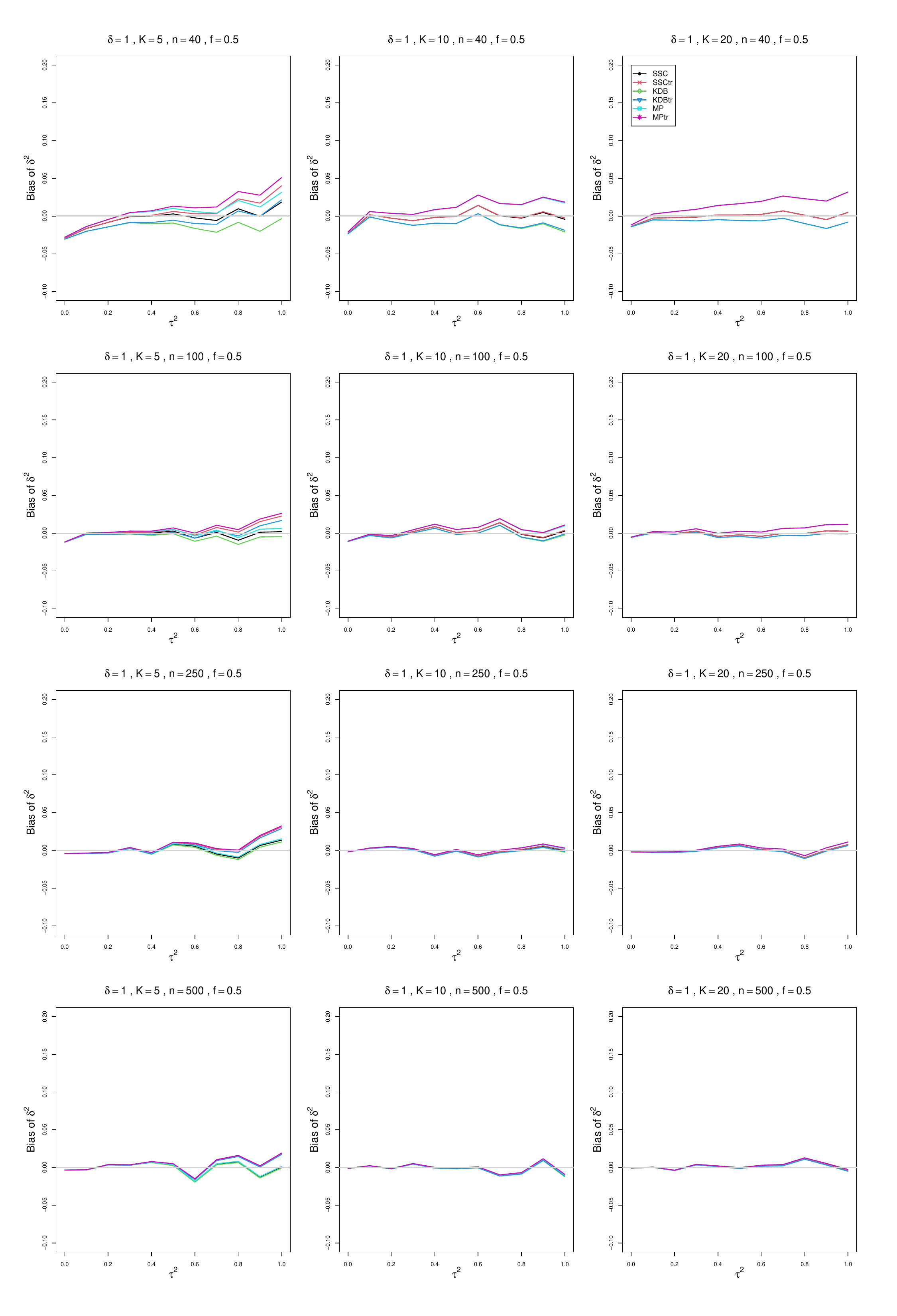}
	\caption{Bias of point estimators of $\delta^2$ based on  MP, KD  and  SMC estimators of $\tau^2$ and their truncated versions vs $\tau^2$, for equal sample sizes $n = 40, \;100,\;250,\;500$, $K=5,\;10$ and $20$, $\delta = 1$ and  $f = 0.5$. }
	\label{PlotBiasOfDelta1Ksmall_equal_sample_sizes.pdf}
\end{figure}

\begin{figure}[ht]
	\centering
	\includegraphics[scale=0.33]{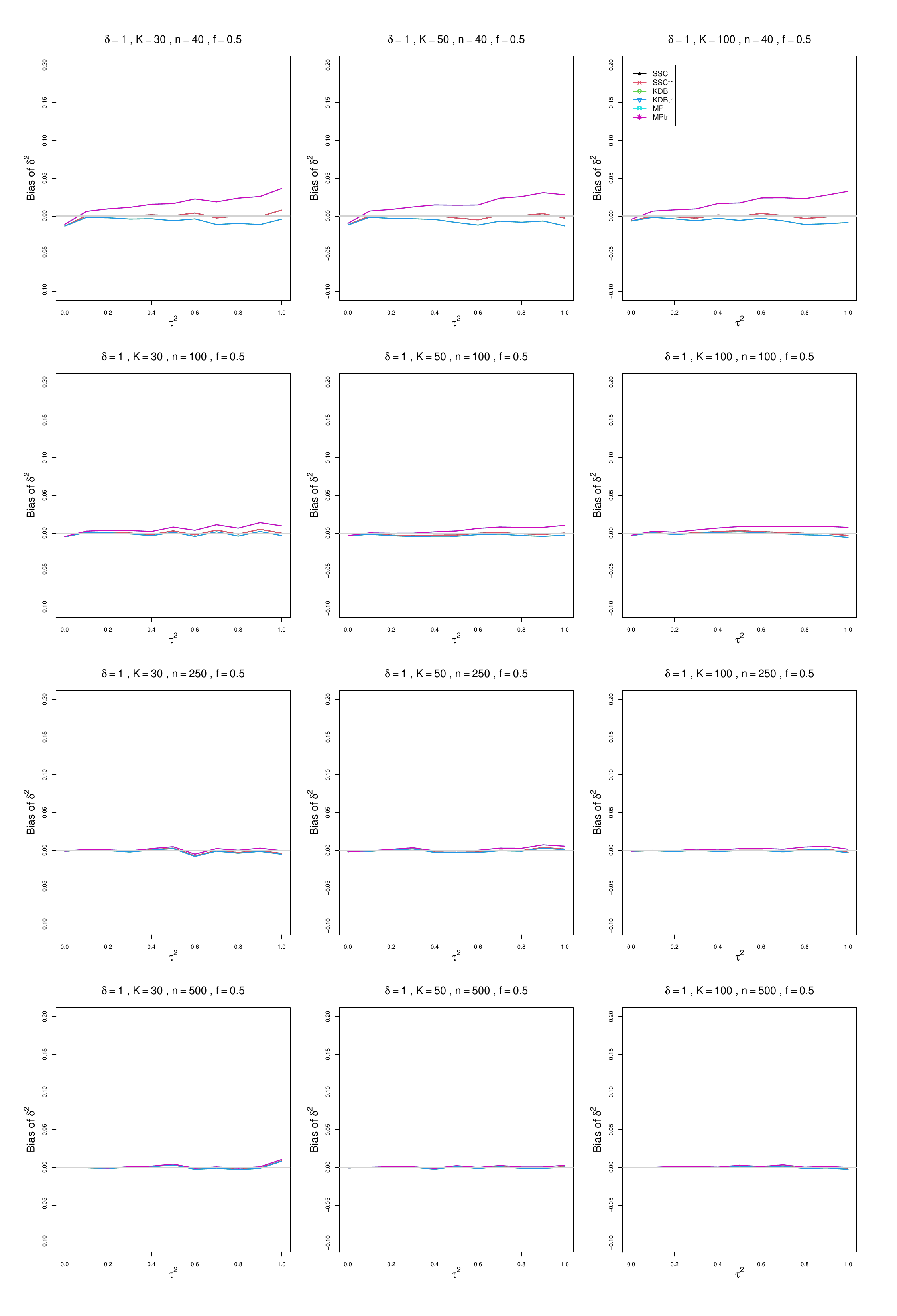}
	\caption{Bias of point estimators of $\delta^2$ based on  MP, KD  and  SMC estimators of $\tau^2$ and their truncated versions vs $\tau^2$, for equal sample sizes $n = 40, \;100,\;250,\;500$, $K=30,\;50$ and $100$, $\delta = 1$ and  $f = 0.5$. }
	\label{PlotBiasOfDelta1Kbig_equal_sample_sizes.pdf}
\end{figure}

\begin{figure}[ht]
	\centering
	\includegraphics[scale=0.33]{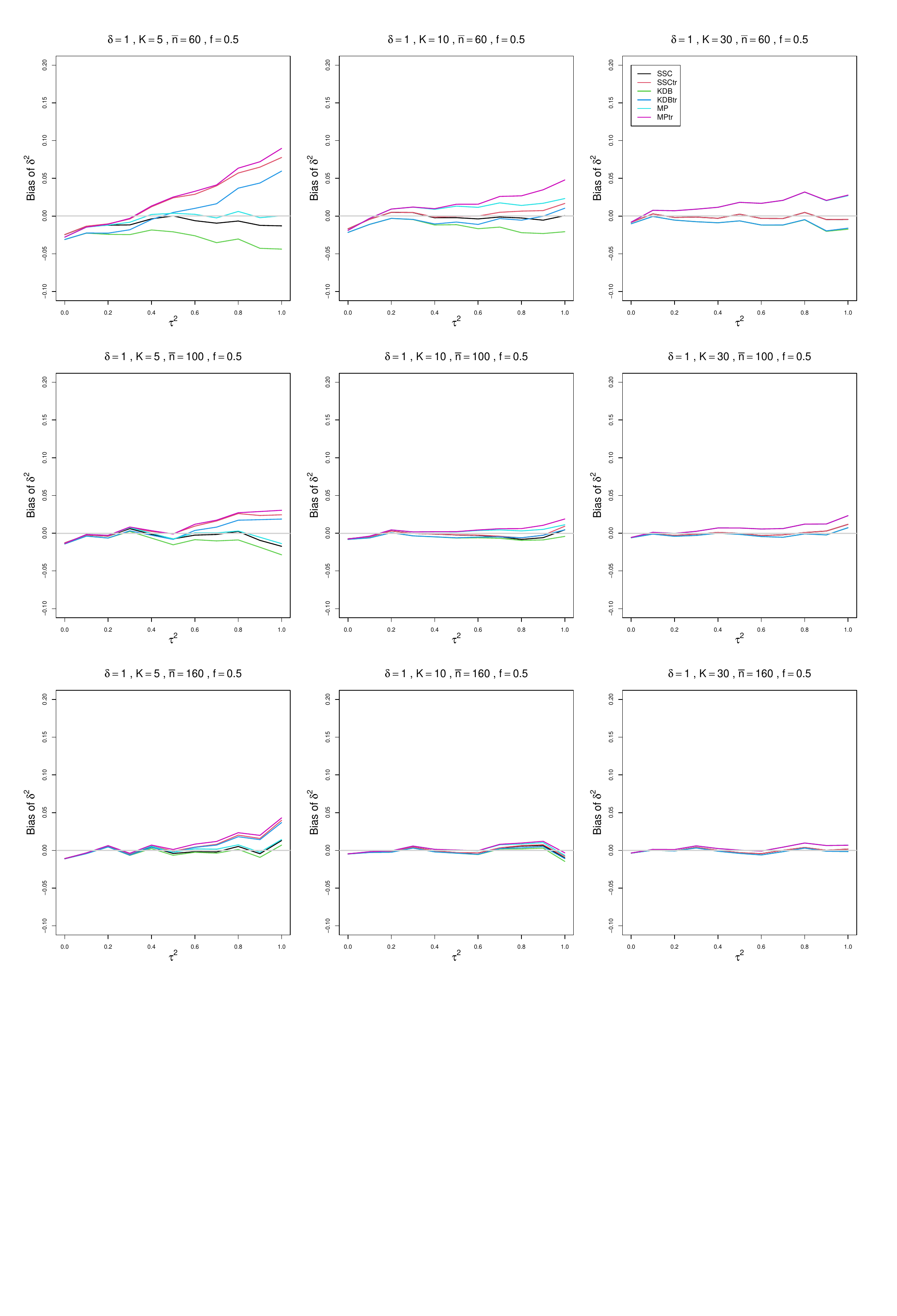}
	\caption{Bias of point estimators of $\delta^2$ based on  MP, KD  and  SMC estimators of $\tau^2$ and their truncated versions vs $\tau^2$, for unequal sample sizes $\bar{n} = 60, \;100,\;160$, $K=5,\;10$ and $30$, $\delta = 1$ and  $f = 0.5$. }
	\label{PlotBiasOfDelta1Kbig_unequal_sample_sizes.pdf}
\end{figure}
\begin{figure}[ht]
	\centering
	\includegraphics[scale=0.33]{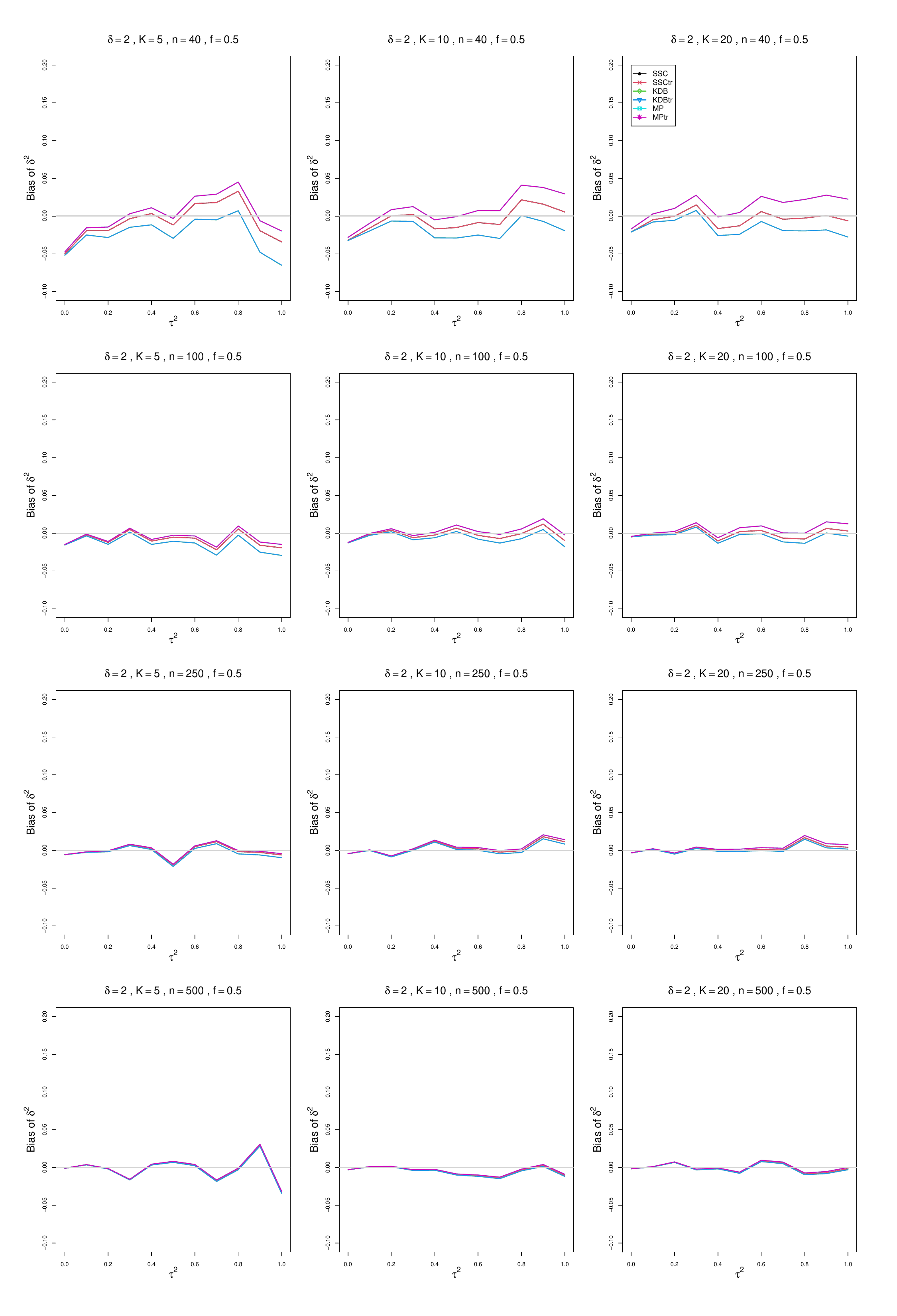}
	\caption{Bias of point estimators of $\delta^2$ based on  MP, KD  and  SMC estimators of $\tau^2$ and their truncated versions vs $\tau^2$, for equal sample sizes $n = 40, \;100,\;250,\;500$, $K=5,\;10$ and $20$, $\delta = 2$ and  $f = 0.5$. }
	\label{PlotBiasOfDelta2Ksmall_equal_sample_sizes.pdf}
\end{figure}

\begin{figure}[ht]
	\centering
	\includegraphics[scale=0.33]{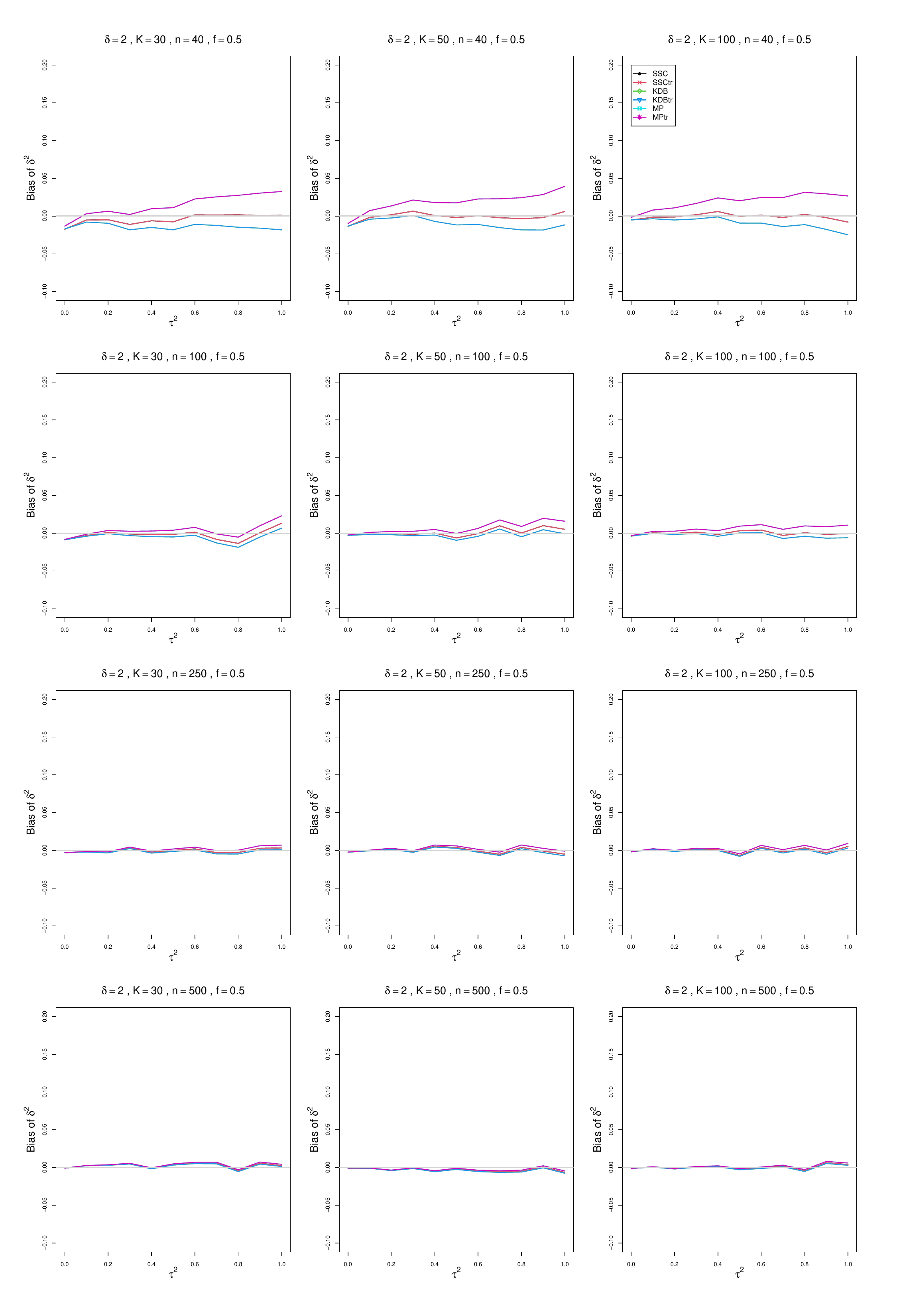}
	\caption{Bias of point estimators of $\delta^2$ based on  MP, KD  and  SMC estimators of $\tau^2$ and their truncated versions vs $\tau^2$, for equal sample sizes $n = 40, \;100,\;250,\;500$, $K=30,\;50$ and $100$, $\delta = 2$ and  $f = 0.5$. }
	\label{PlotBiasOfDelta2Kbig_equal_sample_sizes.pdf}
\end{figure}

\begin{figure}[ht]
	\centering
	\includegraphics[scale=0.33]{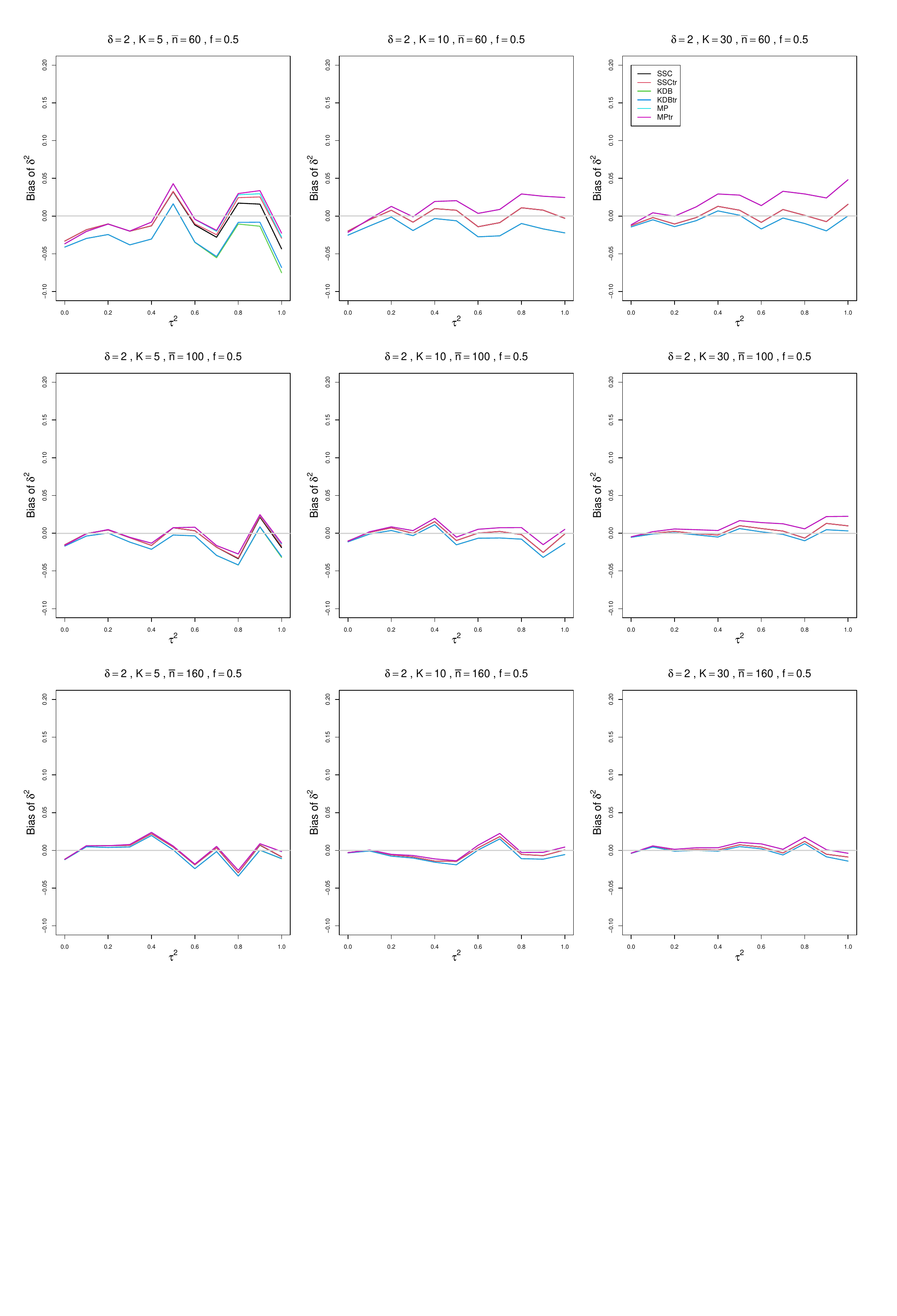}
	\caption{Bias of point estimators of $\delta^2$ based on  MP, KD  and  SMC estimators of $\tau^2$ and their truncated versions vs $\tau^2$, for unequal sample sizes $\bar{n} = 60, \;100,\;160$, $K=5,\;10$ and $30$, $\delta = 2$ and  $f = 0.5$. }
	\label{PlotBiasOfDelta2Kbig_unequal_sample_sizes.pdf}
\end{figure}

\clearpage

\section*{Appendix B: Empirical level of conditional tests of $\delta^2 = 0$ at a 5\% nominal level}

Each figure corresponds to the standardized mean difference $\delta=0$.
The fraction of each study's sample size in the Control arm ($f$) is held constant at 0.5.

For each combination of a value of $n$ (=  40, 100, 250, 500) or  $\bar{n}$ (= 60, 100, 160) and a value of $K$ (= 5, 10, 20 or 30, 50, 100), a panel plots levels of conditional (given $\hat\tau^2$) $\Lambda(\hat\tau^2)$ tests of $\delta^2 = 0$ versus $\tau^2$ (= 0(0.1)1).\\
The  tests  are
\begin{itemize}
\item KDB,  conditional test given $\tau^2_{KDB}$ (Kulinskaya-Dollinger-Bj{\o}rkest{\o}l method), $\chi^2_K$ approximation
\item MP, conditional test given $\tau^2_{MP}$ (Mandel-Paule method), $\chi^2_K$ approximation
\item SSC,  conditional test given $\tau^2_{SSC}$, $\chi^2_K$ approximation
\item KDB\_b,  conditional test given $\tau^2_{KDB}$ (Kulinskaya-Dollinger-Bj{\o}rkest{\o}l method),
bootstrap p-value, B=100000
\item MP\_b, conditional test given $\tau^2_{MP}$ (Mandel-Paule method), bootstrap p-value, B=100000
\item SSC\_b,  conditional test given $\tau^2_{SSC}$, bootstrap p-value, B=100000
\item $\tau^2$, a comparator test using known $\tau^2$ value,  bootstrap p-value, B=100000
\end{itemize}

\clearpage

\setcounter{figure}{0}
\setcounter{section}{0}
\renewcommand{\thefigure}{B.\arabic{figure}}

\begin{figure}[ht]
	\centering
	\includegraphics[scale=0.33]{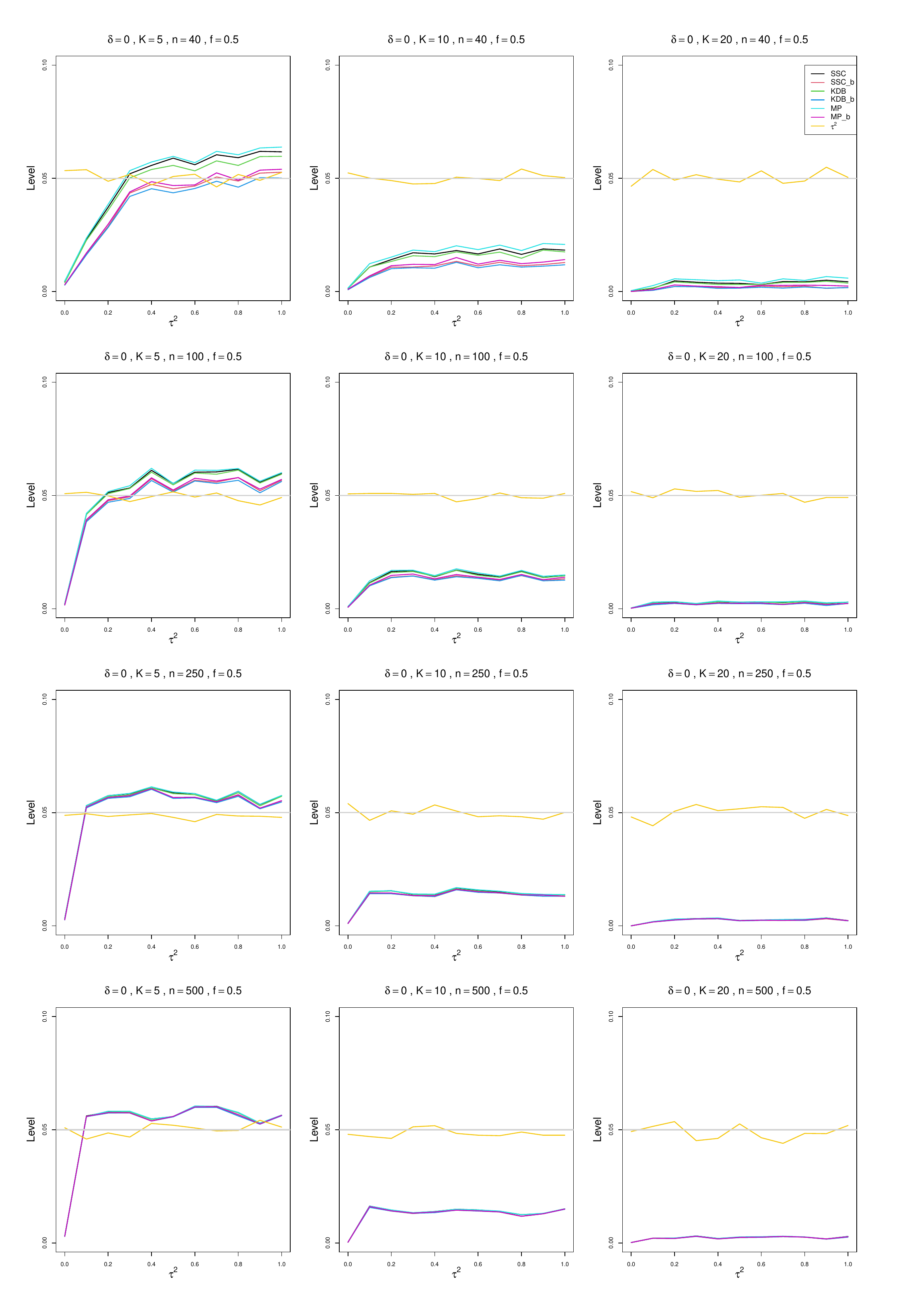}
	\caption{Empirical level of conditional tests of $\delta^2=0$ and  a comparator test with known $\tau^2$ at 5\% nominal level:   MP, KDB, SMC tests, using $\chi^2_K$ approximation or a bootstrap p-value vs $\tau^2$,   for equal sample sizes $n = 40, \;100,\;250,\;500$, $K=5,\;10$ and $20$, $\delta = 0$ and  $f = 0.5$. }
	\label{PPlotLevelsOfLambda2 equal_sample_sizes_k5_10_20}
\end{figure}

\begin{figure}[ht]
	\centering
	\includegraphics[scale=0.33]{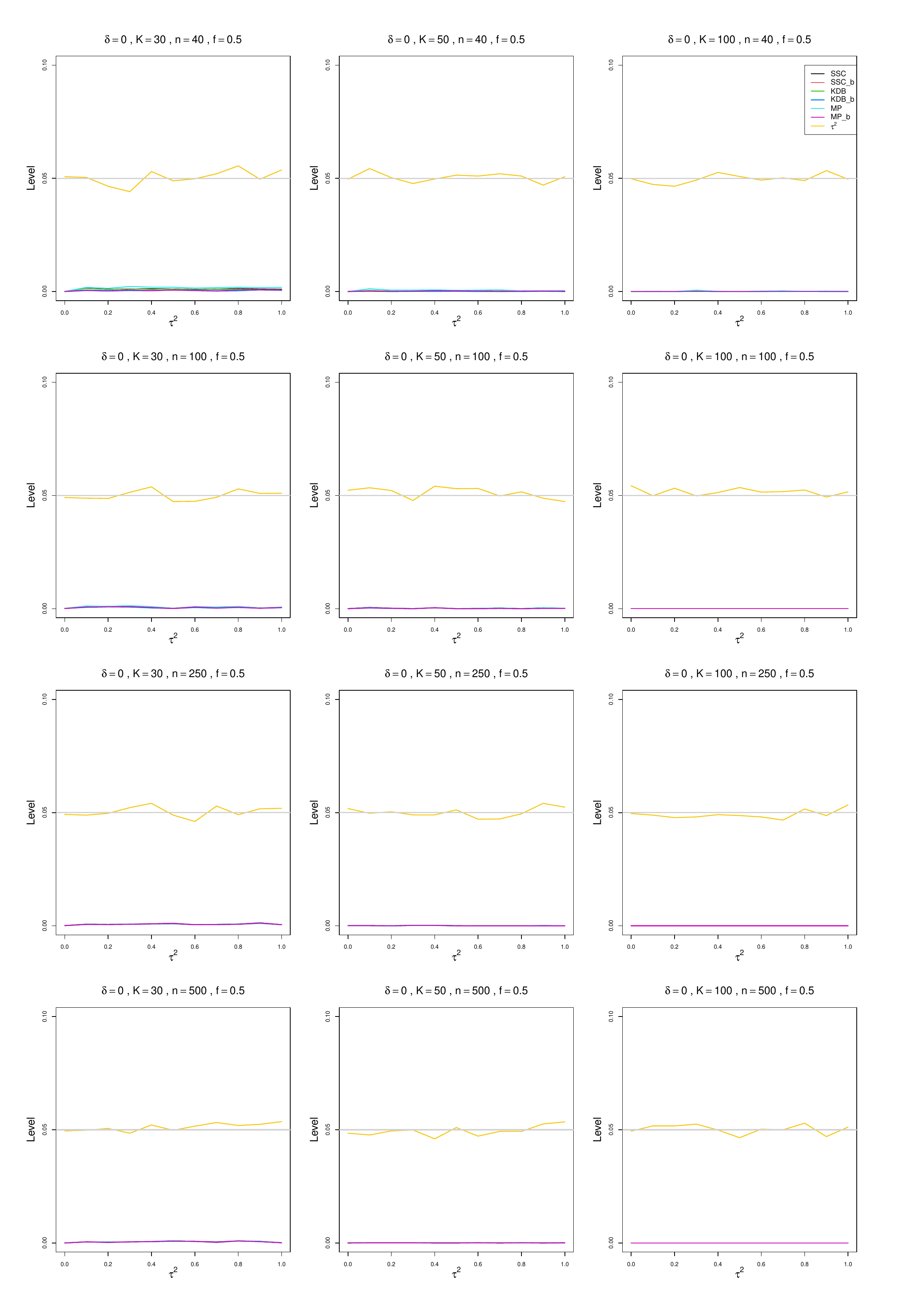}
	\caption{Empirical level of conditional tests of $\delta^2=0$ and  a comparator test with known $\tau^2$ at 5\% nominal level:   MP, KDB, SMC tests, using $\chi^2_K$ approximation or a bootstrap p-value vs $\tau^2$,   for equal sample sizes $n = 40, \;100,\;250,\;500$, $K=30,\;50$ and $100$, $\delta = 0$ and  $f = 0.5$. }
	\label{PlotLevelsOfLambda2 equal_sample_sizes_k30_50_100}
\end{figure}

\begin{figure}[ht]
	\centering
	\includegraphics[scale=0.33]{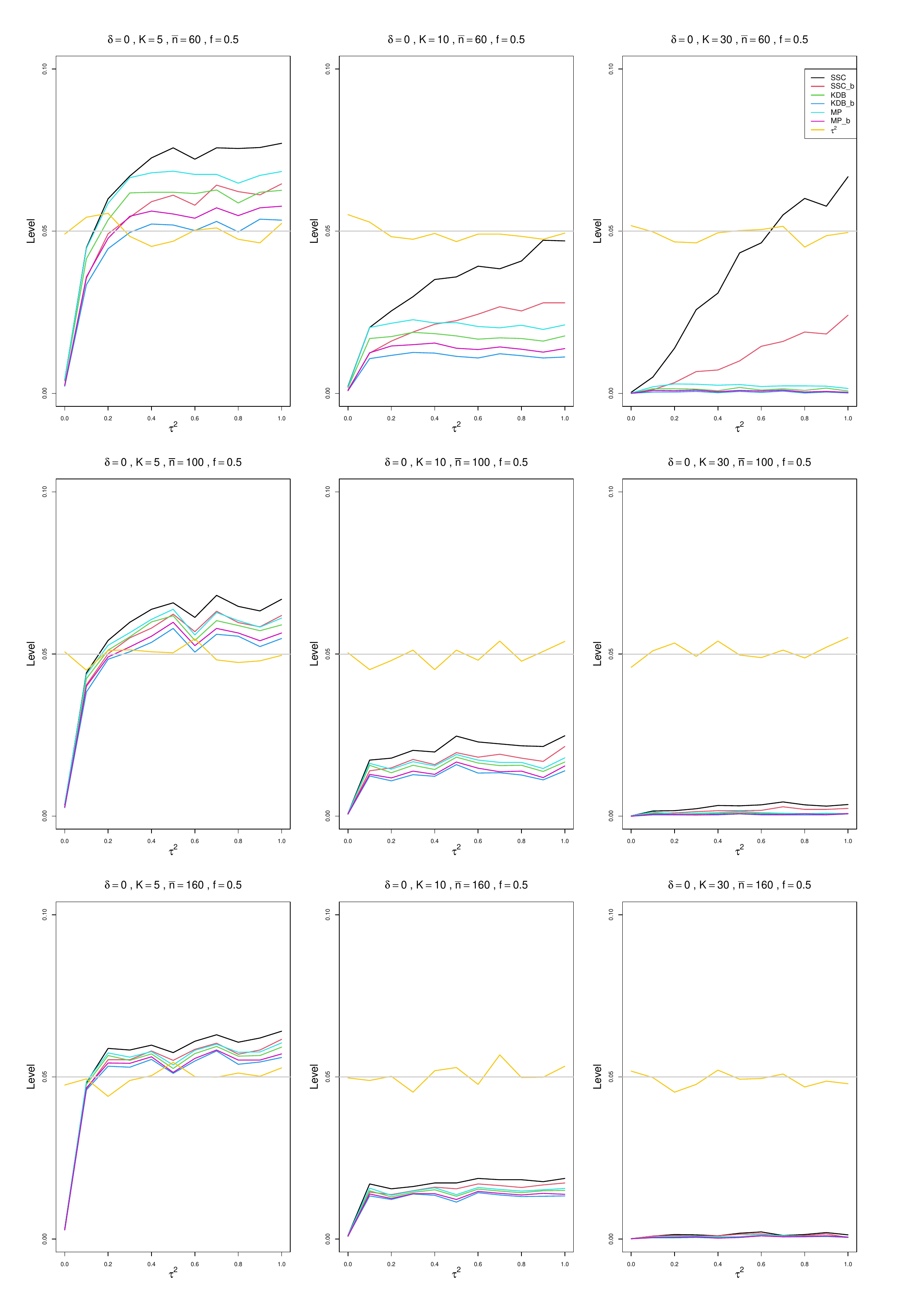}
	\caption{Empirical level of conditional tests of $\delta^2=0$ and  a comparator test with known $\tau^2$ at 5\% nominal level:   MP, KDB, SMC tests, using $\chi^2_K$ approximation or a bootstrap p-value vs $\tau^2$,   for unequal sample sizes $\bar{n} = 60, \;100,\;160$, $K=5,\;10$ and $30$, $\delta = 0$ and  $f = 0.5$. }
	\label{PlotLevelsOfLambda2_0_ASMD_unequal_sample_sizes}
\end{figure}

\clearpage

\section*{Appendix C: Coverage of 95\% confidence intervals for $\delta^2$}

Each figure corresponds to a value of the standardized mean difference $\delta$ (=0, 0.2, 0.5, 1, 2).
The fraction of each study's sample size in the Control arm ($f$) is held constant at 0.5.

For each combination of a value of $n \in$ \{40, 100, 250, 500\} or  $\bar{n} \in$ \{60, 100, 160\} and a value of $K \in$ \{5, 10, 20 or 30, 50, 100\}, a panel plots coverage of $\delta^2$ versus $\tau^2$ (= 0(0.1)1).\\
The interval  estimators of $\delta^2$ are
\begin{itemize}
\item KDB (Kulinskaya-Dollinger-Bj{\o}rkest{\o}l) method, inverse-variance weights, normal quantiles, based on the unsigned SMD values
\item MP (Mandel-Paule) method, inverse-variance weights, normal quantiles, based on the signed SMD values
\item SSC method, effective-sample-size weights, normal quantiles, based on the signed SMD values
\item SSC\_t method, effective-sample-size weights, $t_{K-1}$ quantiles, based on the signed SMD values
\item KDB\_c,  conditional interval given $\tau^2_{KDB}$ (Kulinskaya-Dollinger-Bj{\o}rkest{\o}l) method
\item MP\_c, conditional interval given $\tau^2_{MP}$ (Mandel-Paule method)
\item SSC\_c conditional interval given $\tau^2_{SSC}$
\end{itemize}
Unconditional intervals have two versions: a na\"ive version using confidence intervals $(0, \max(L^2,U^2))$ or $(\min(L^2,U^2), \max(L^2,U^2))$ where $(L,U)$ is a CI for $\delta$, (dashed lines) and a corrected version, denoted by * (straight lines). See text for details.

\clearpage

\setcounter{figure}{0}
\setcounter{section}{0}
\renewcommand{\thefigure}{C.\arabic{figure}}

\begin{figure}[ht]
	\centering
	\includegraphics[scale=0.33]{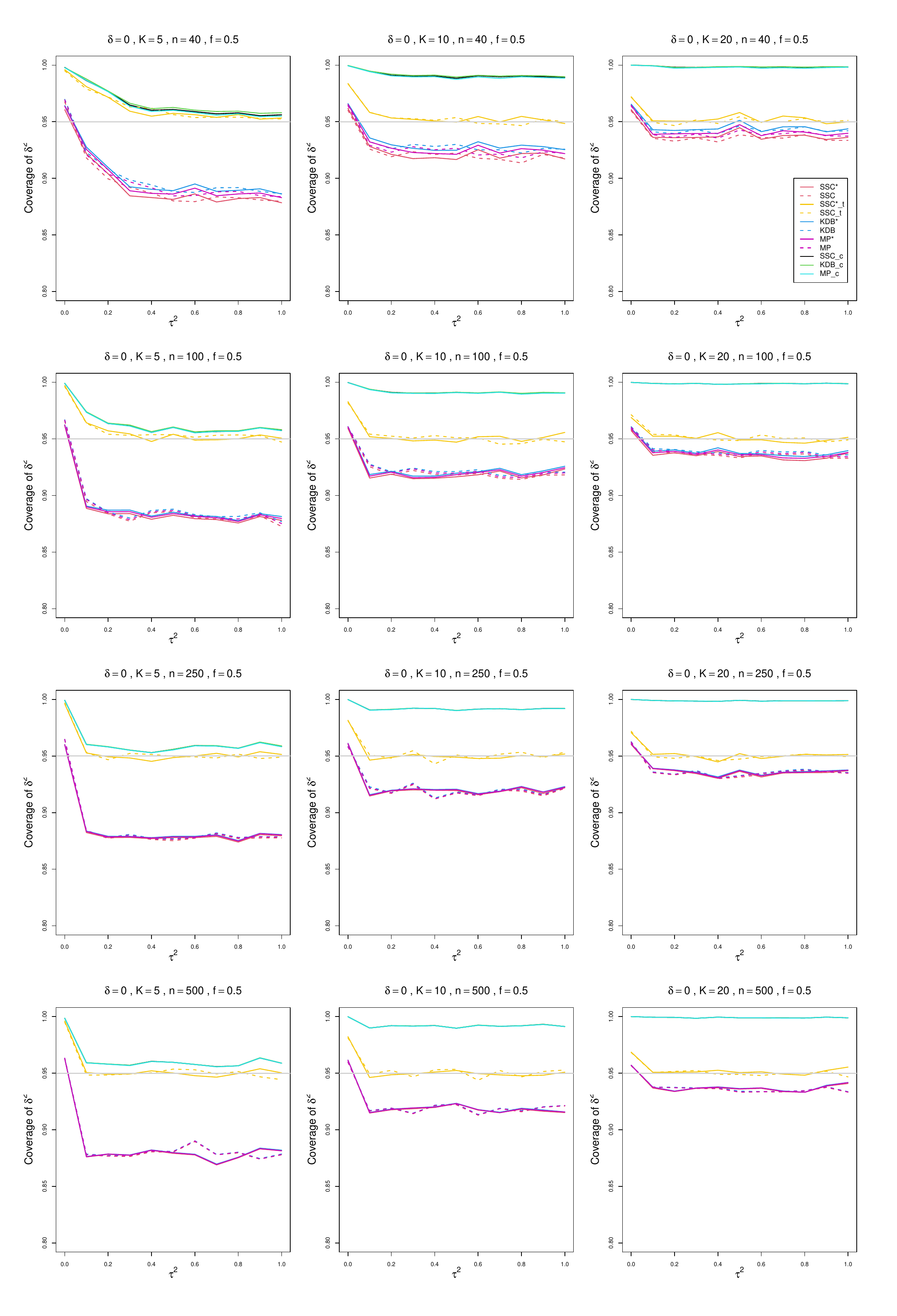}
	\caption{Coverage  of confidence intervals of $\delta^2$ at 95\% nominal level:   MP, KD, SMC and SMC-t intervals, na\"ive and corrected(*), based on the signed SMD values and conditional intervals  (MP\_c, KD\_c  and  SMC\_c) vs $\tau^2$, for equal sample sizes $n = 40, \;100,\;250,\;500$, $K=5,\;10$ and $20$, $\delta = 0$ and  $f = 0.5$. }
	\label{PlotCovOfDelta0Ksmall_equal_sample_sizes.pdf}
\end{figure}

\begin{figure}[ht]
	\centering
	\includegraphics[scale=0.33]{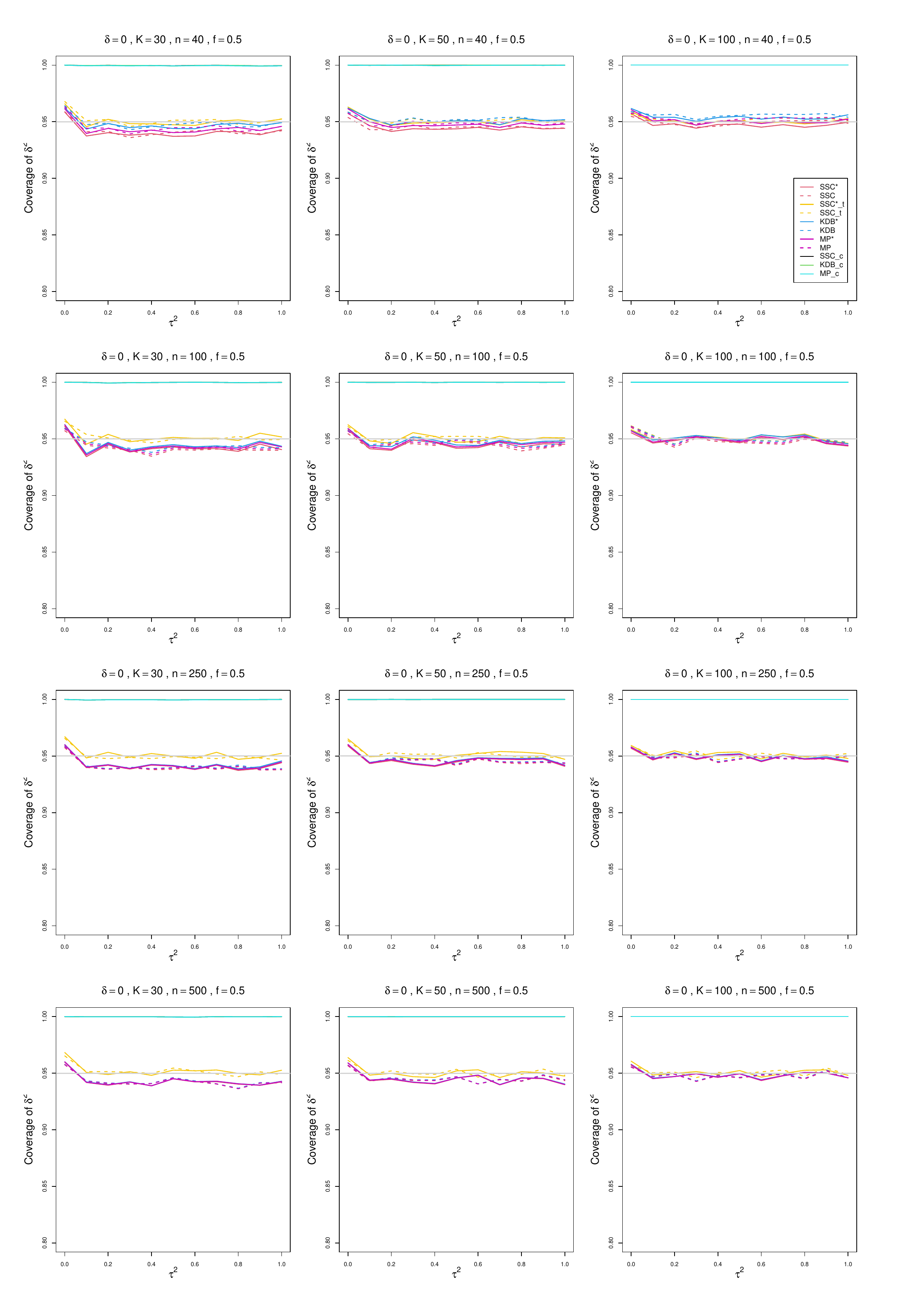}
	\caption{Coverage  of confidence intervals of $\delta^2$ at 95\% nominal level:   MP, KD, SMC and SMC-t intervals, na\"ive and corrected(*),  based on the signed SMD values and conditional intervals  (MP\_c, KD\_c  and  SMC\_c) vs $\tau^2$, for equal sample sizes $n = 100$, $K=30,\;50$ and $100$, $\delta = 0$  and  $f = 0.5$ f }
	\label{PlotcovOfDelta0Kbig_equal_sample_sizes.pdf}
\end{figure}

\begin{figure}[ht]
	\centering
	\includegraphics[scale=0.33]{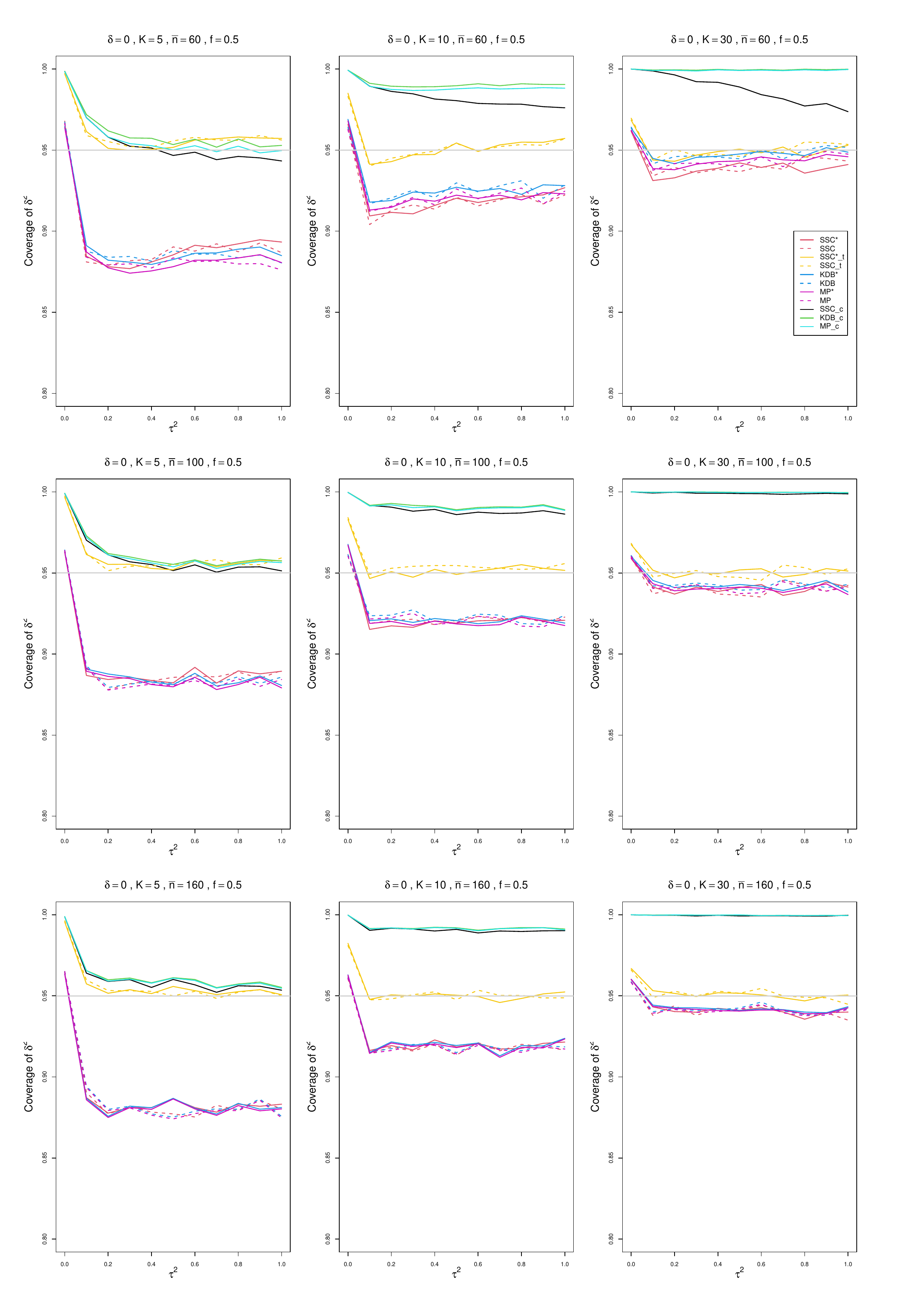}
	\caption{Coverage  of confidence intervals of $\delta^2$ at 95\% nominal level:   MP, KD, SMC and SMC-t intervals, na\"ive and corrected(*),  based on the signed SMD values and conditional intervals  (MP\_c, KD\_c  and  SMC\_c) vs $\tau^2$, for unequal sample sizes $\bar{n} = 60, \;100,\;160$, $K=5,\;10$ and $30$, $\delta = 0$ and  $f = 0.5$. }
	\label{PlotCovOfDelta0Ksmall_unequal_sample_sizes.pdf}
\end{figure}

\begin{figure}[ht]
	\centering
	\includegraphics[scale=0.33]{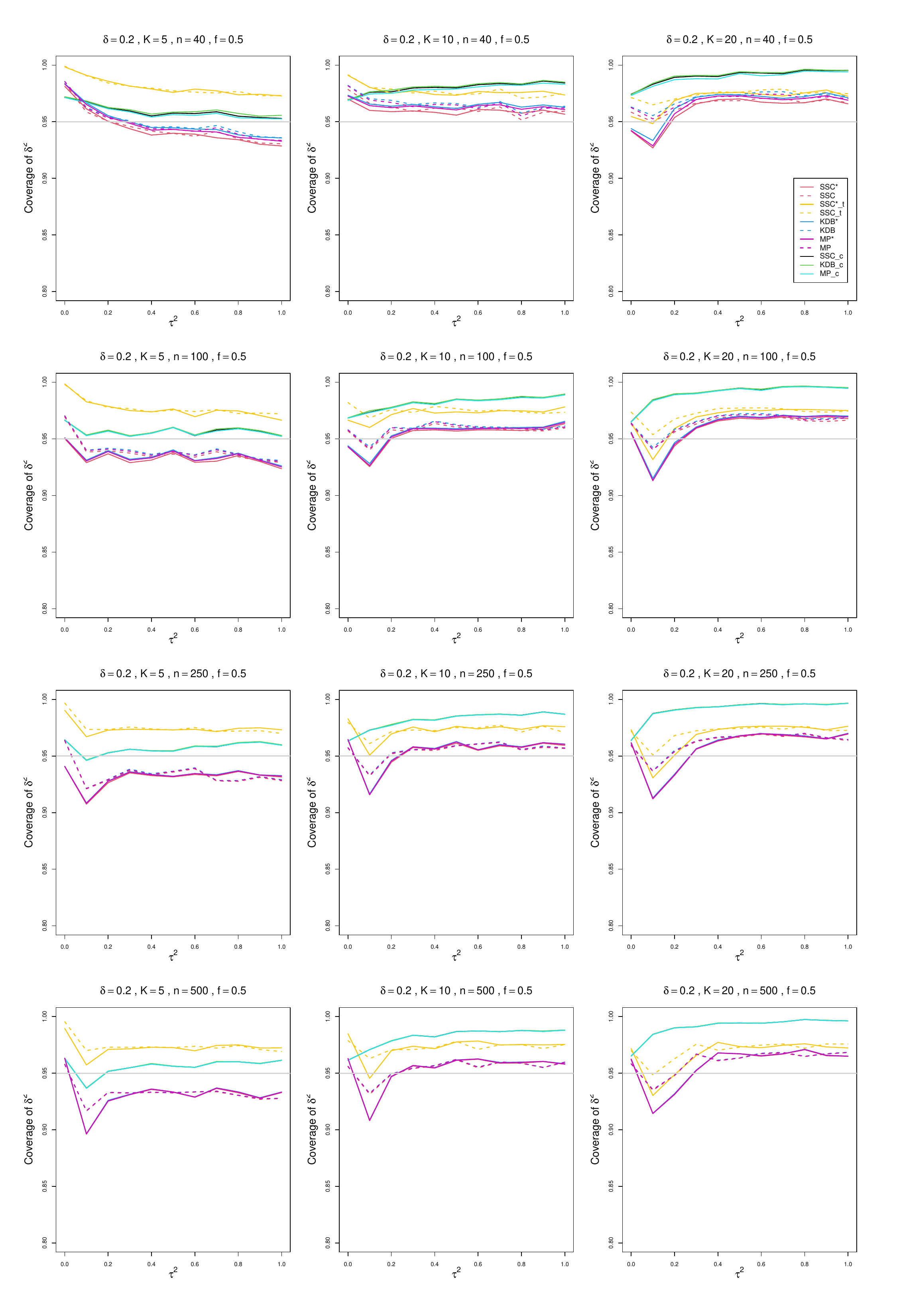}
	\caption{Coverage  of confidence intervals of $\delta^2$ at 95\% nominal level:   MP, KD, SMC and SMC-t intervals, na\"ive and corrected(*),  based on the signed SMD values and conditional intervals  (MP\_c, KD\_c  and  SMC\_c) vs $\tau^2$, for equal sample sizes $n = 40, \;100,\;250,\;500$, $K=5,\;10$ and $20$, $\delta = 0.2$ and  $f = 0.5$. }
	\label{PlotCovOfDelta02Ksmall_equal_sample_sizes.pdf}
\end{figure}

\begin{figure}[ht]
	\centering
	\includegraphics[scale=0.33]{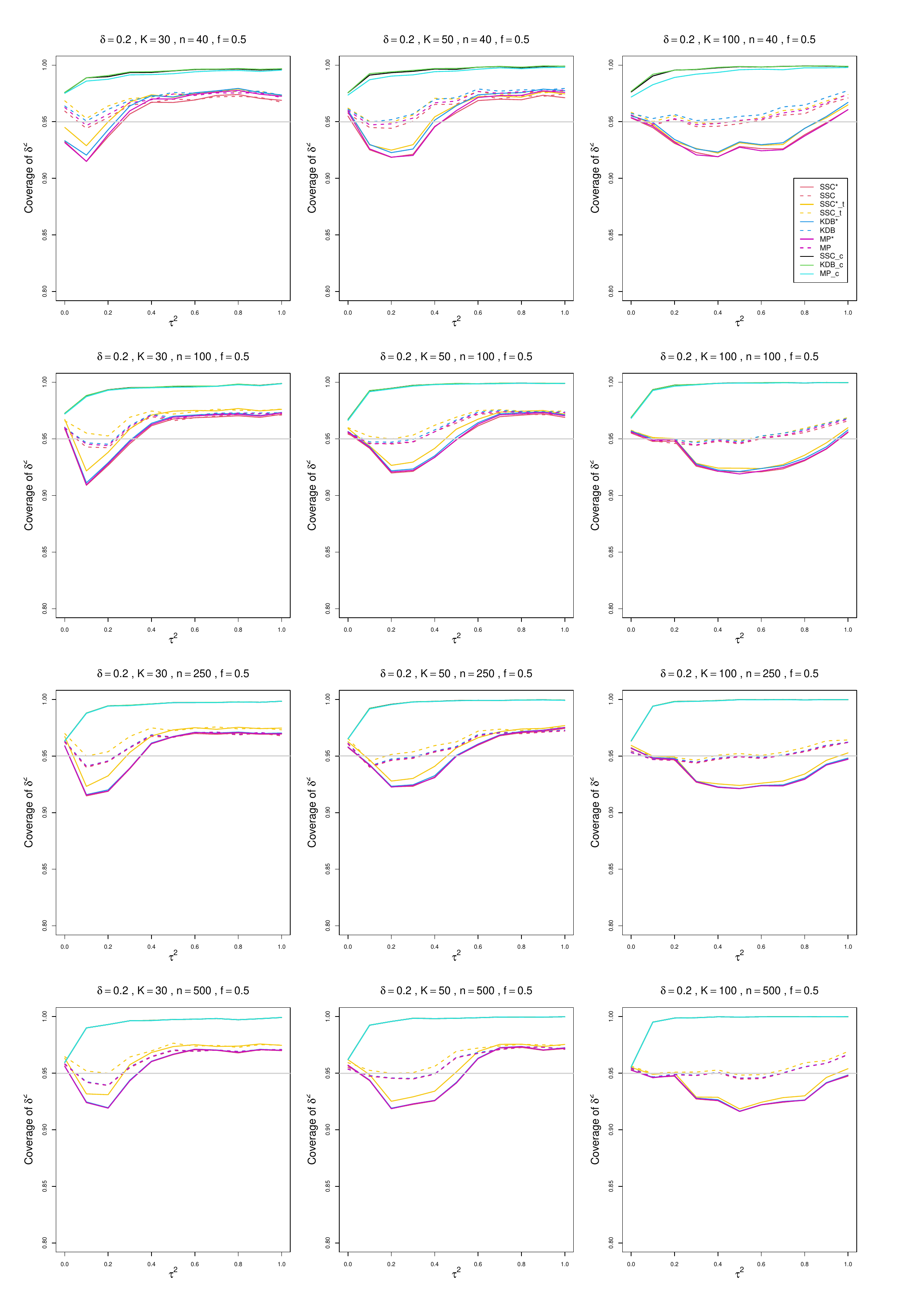}
	\caption{Coverage  of confidence intervals of $\delta^2$ at 95\% nominal level:   MP, KD, SMC and SMC-t intervals, na\"ive and corrected(*),  based on the signed SMD values and conditional intervals  (MP\_c, KD\_c  and  SMC\_c) vs $\tau^2$, for equal sample sizes $n = 100$, $K=30,\;50$ and $100$, $\delta = 0.2$  and  $f = 0.5$ f }
	\label{PlotCovOfDelta02Kbig_equal_sample_sizes.pdf}
\end{figure}

\begin{figure}[ht]
	\centering
	\includegraphics[scale=0.33]{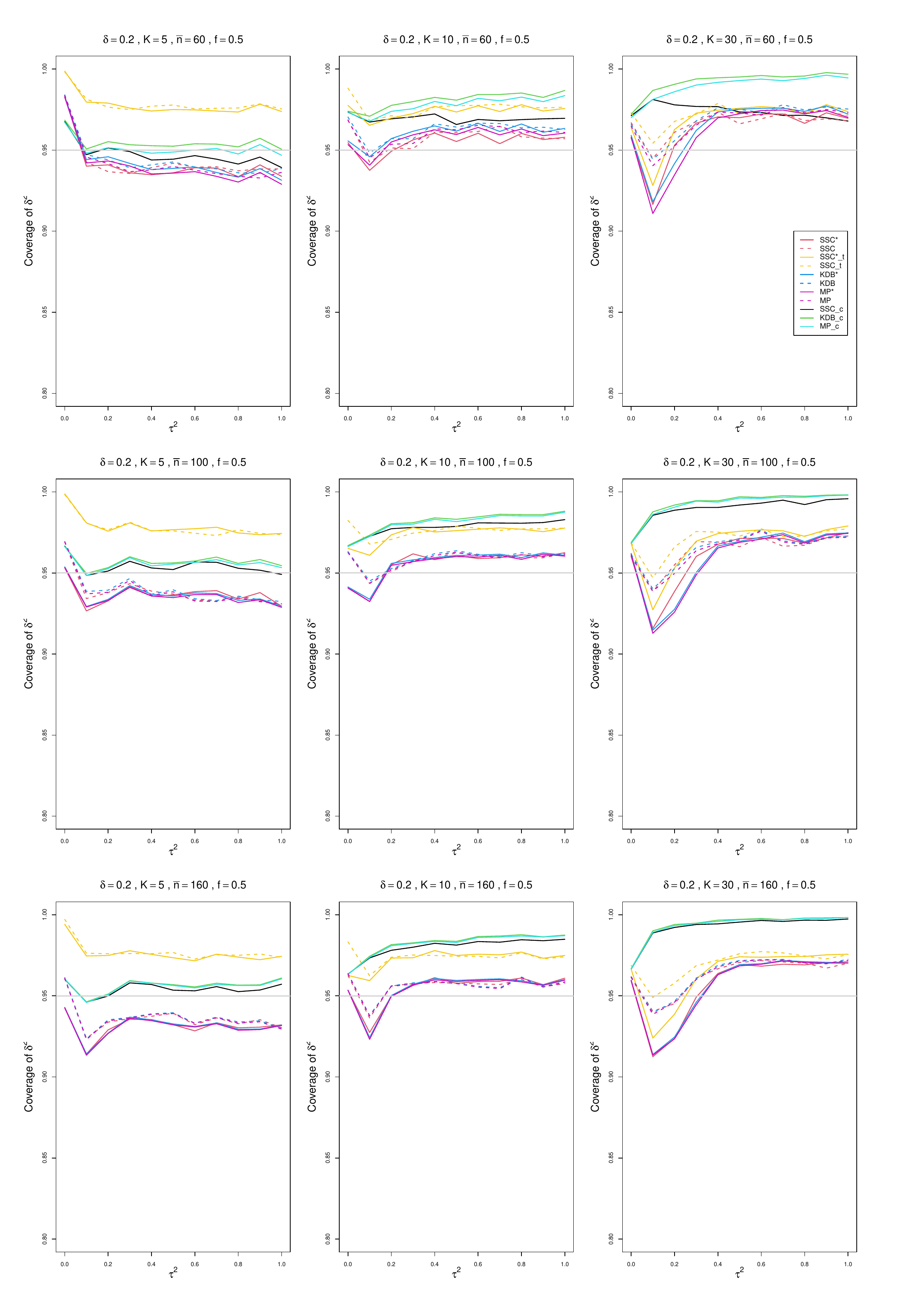}
	\caption{Coverage  of confidence intervals of $\delta^2$ at 95\% nominal level:   MP, KD, SMC and SMC-t intervals, na\"ive and corrected(*),  based on the signed SMD values and conditional intervals  (MP\_c, KD\_c  and  SMC\_c) vs $\tau^2$, for unequal sample sizes $\bar{n} = 60, \;100,\;160$, $K=5,\;10$ and $30$, $\delta = 0.2$ and  $f = 0.5$. }
	\label{PlotCovOfDelta02Ksmall_unequal_sample_sizes.pdf}
\end{figure}

\begin{figure}[ht]
	\centering
	\includegraphics[scale=0.33]{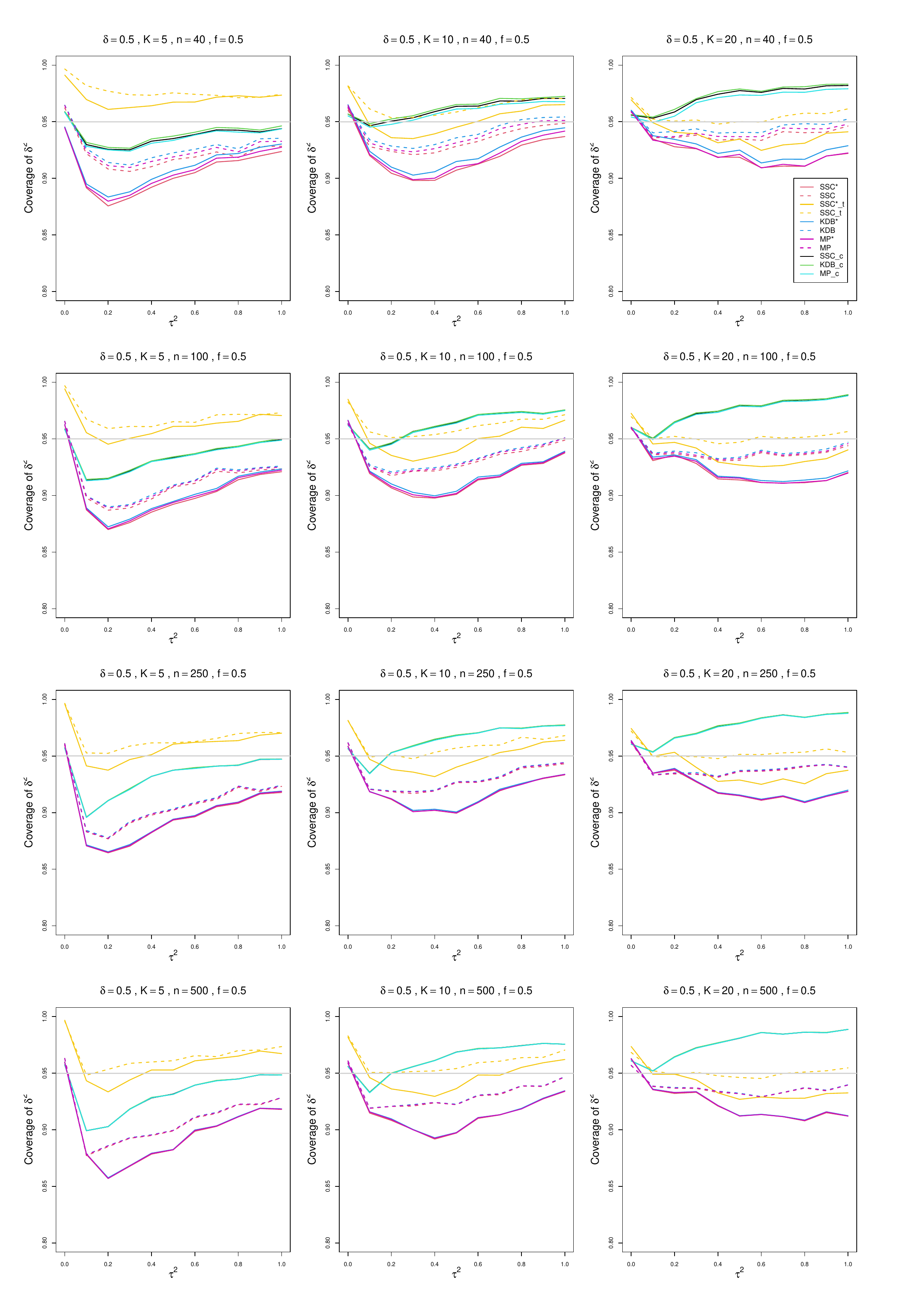}
	\caption{Coverage  of confidence intervals of $\delta^2$ at 95\% nominal level:   MP, KD, SMC and SMC-t intervals, na\"ive and corrected(*),  based on the signed SMD values and conditional intervals  (MP\_c, KD\_c  and  SMC\_c) vs $\tau^2$, for equal sample sizes $n = 40, \;100,\;250,\;500$, $K=5,\;10$ and $20$, $\delta = 0.5$ and  $f = 0.5$. }
	\label{PlotCovOfDelta05Ksmall_equal_sample_sizes.pdf}
\end{figure}

\begin{figure}[ht]
	\centering
	\includegraphics[scale=0.33]{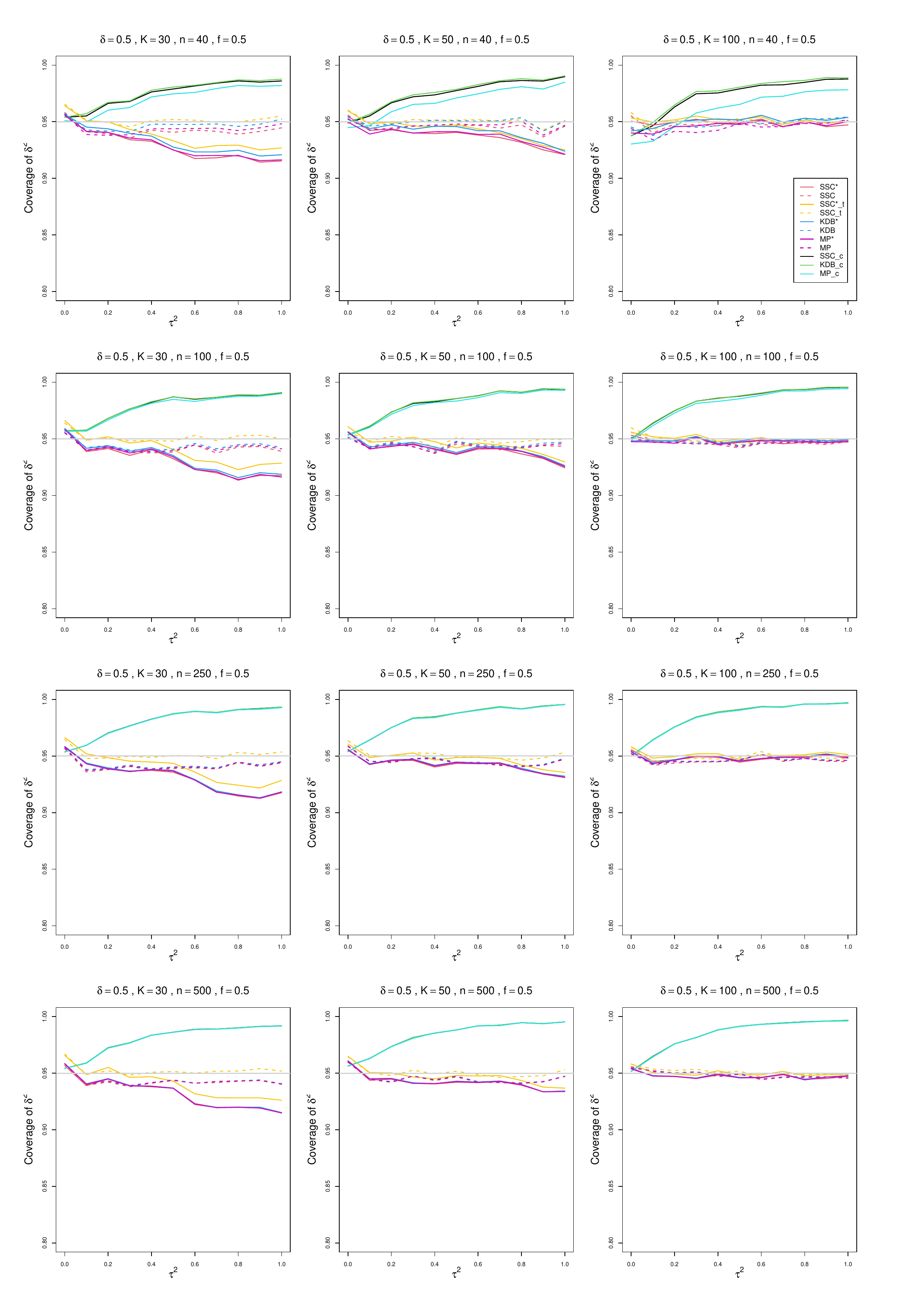}
	\caption{Coverage  of confidence intervals of $\delta^2$ at 95\% nominal level:   MP, KD, SMC and SMC-t intervals, na\"ive and corrected(*),  based on the signed SMD values and conditional intervals  (MP\_c, KD\_c  and  SMC\_c) vs $\tau^2$, for equal sample sizes $n = 100$, $K=30,\;50$ and $100$, $\delta = 0.5$  and  $f = 0.5$ f }
	\label{PlotCovOfDelta05Kbig_equal_sample_sizes.pdf}
\end{figure}

\begin{figure}[ht]
	\centering
	\includegraphics[scale=0.33]{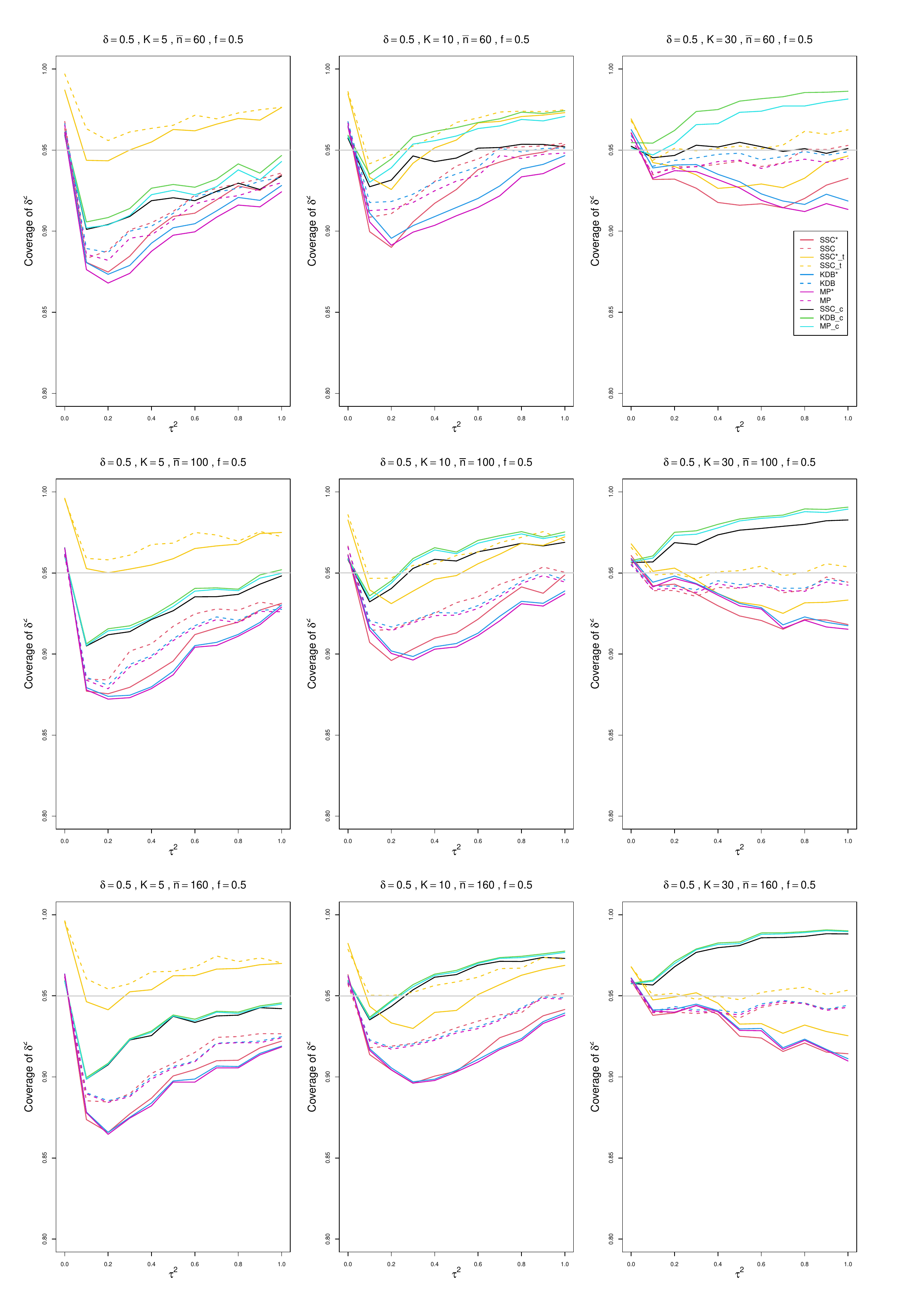}
	\caption{Coverage  of confidence intervals of $\delta^2$ at 95\% nominal level:   MP, KD, SMC and SMC-t intervals, na\"ive and corrected(*),  based on the signed SMD values and conditional intervals  (MP\_c, KD\_c  and  SMC\_c) vs $\tau^2$, for unequal sample sizes $\bar{n} = 60, \;100,\;160$, $K=5,\;10$ and $30$, $\delta = 0.5$ and  $f = 0.5$. }
	\label{PlotCovOfDelta05Ksmall_unequal_sample_sizes.pdf}
\end{figure}

\begin{figure}[ht]
	\centering
	\includegraphics[scale=0.33]{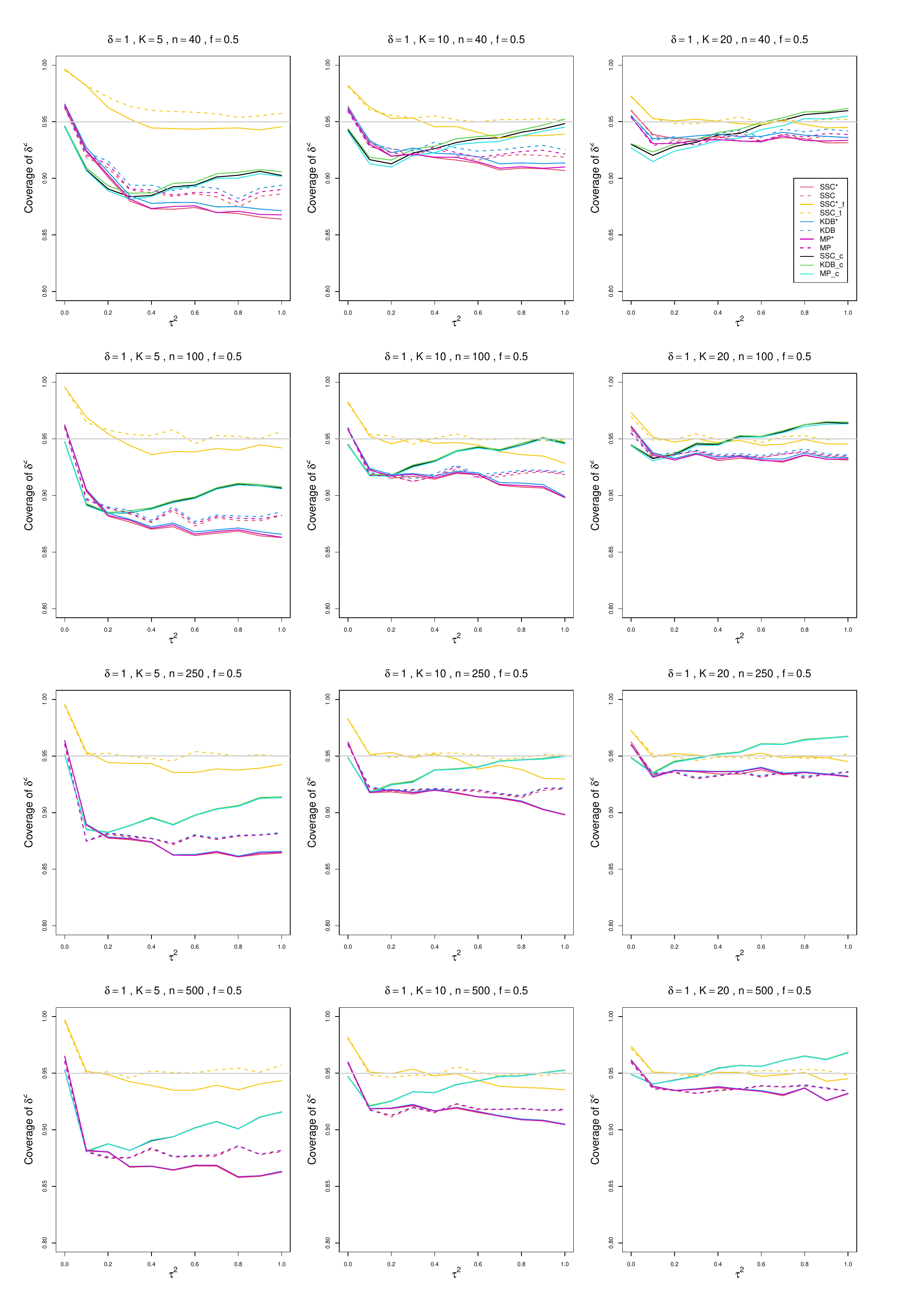}
	\caption{Coverage  of confidence intervals of $\delta^2$ at 95\% nominal level:   MP, KD, SMC and SMC-t intervals, na\"ive and corrected(*),  based on the signed SMD values and conditional intervals  (MP\_c, KD\_c  and  SMC\_c) vs $\tau^2$, for equal sample sizes $n = 40, \;100,\;250,\;500$, $K=5,\;10$ and $20$, $\delta = 1$ and  $f = 0.5$. }
	\label{PlotCovOfDelta1Ksmall_equal_sample_sizes.pdf}
\end{figure}

\begin{figure}[ht]
	\centering
	\includegraphics[scale=0.33]{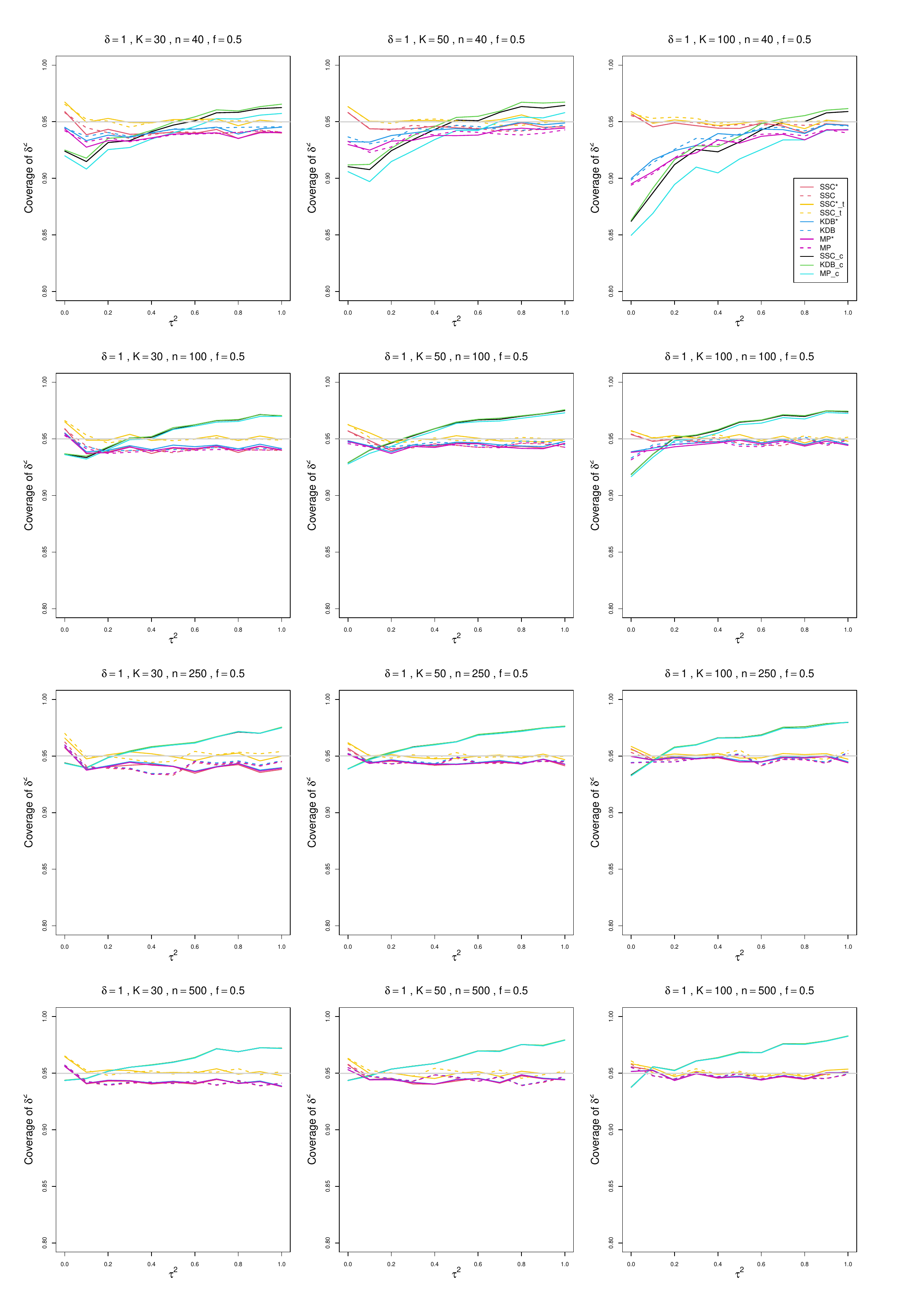}
	\caption{Coverage  of confidence intervals of $\delta^2$ at 95\% nominal level:   MP, KD, SMC and SMC-t intervals, na\"ive and corrected(*),  based on the signed SMD values and conditional intervals  (MP\_c, KD\_c  and  SMC\_c) vs $\tau^2$, for equal sample sizes $n = 100$, $K=30,\;50$ and $100$, $\delta = 1$  and  $f = 0.5$ f }
	\label{PlotCovOfDelta1Kbig_equal_sample_sizes.pdf}
\end{figure}

\begin{figure}[ht]
	\centering
	\includegraphics[scale=0.33]{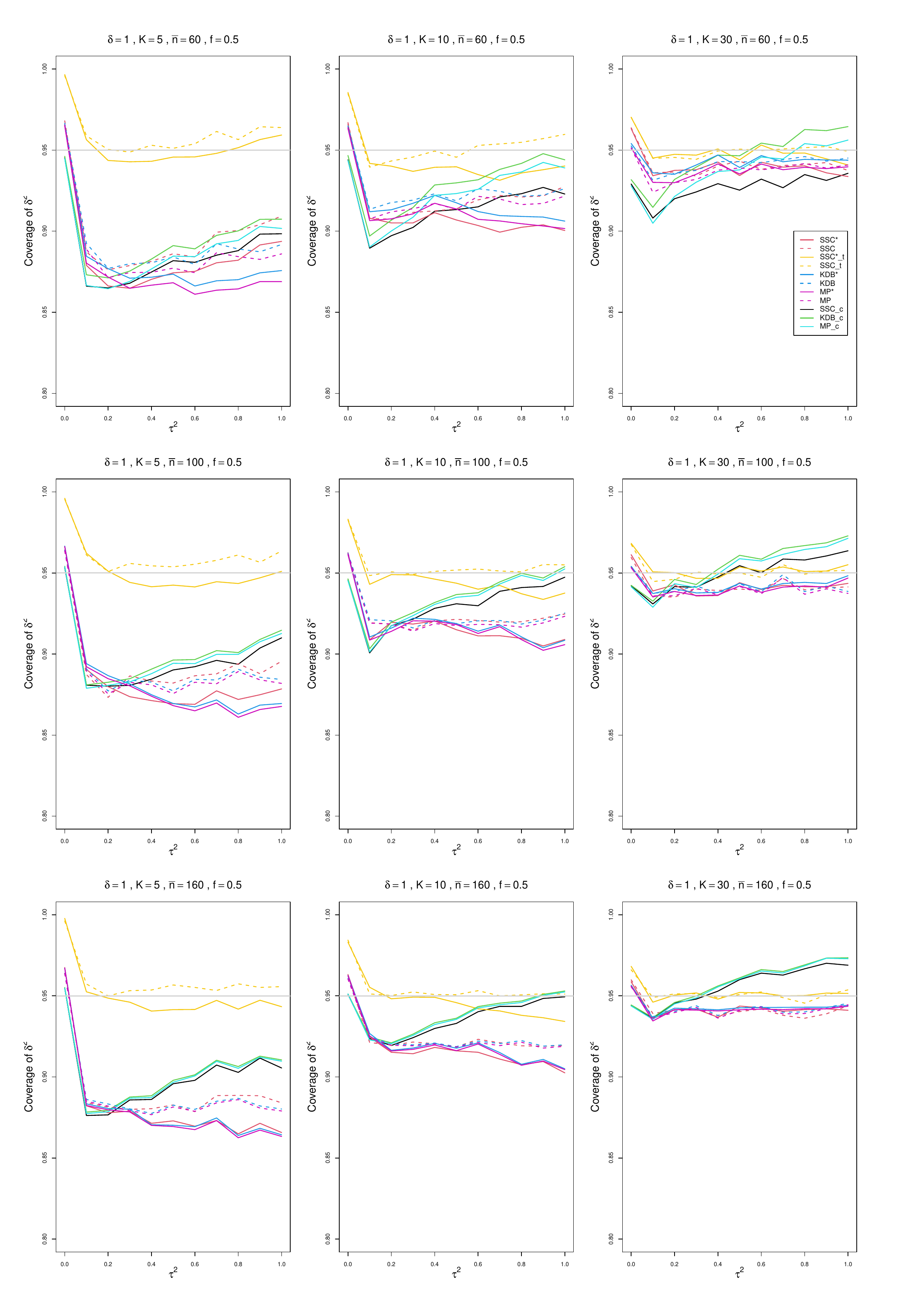}
	\caption{Coverage  of confidence intervals of $\delta^2$ at 95\% nominal level:   MP, KD, SMC and SMC-t intervals, na\"ive and corrected(*),  based on the signed SMD values and conditional intervals  (MP\_c, KD\_c  and  SMC\_c) vs $\tau^2$, for unequal sample sizes $\bar{n} = 60, \;100,\;160$, $K=5,\;10$ and $30$, $\delta = 1$ and  $f = 0.5$. }
	\label{PlotCovOfDelta1Ksmall_unequal_sample_sizes.pdf}
\end{figure}

\begin{figure}[ht]
	\centering
	\includegraphics[scale=0.33]{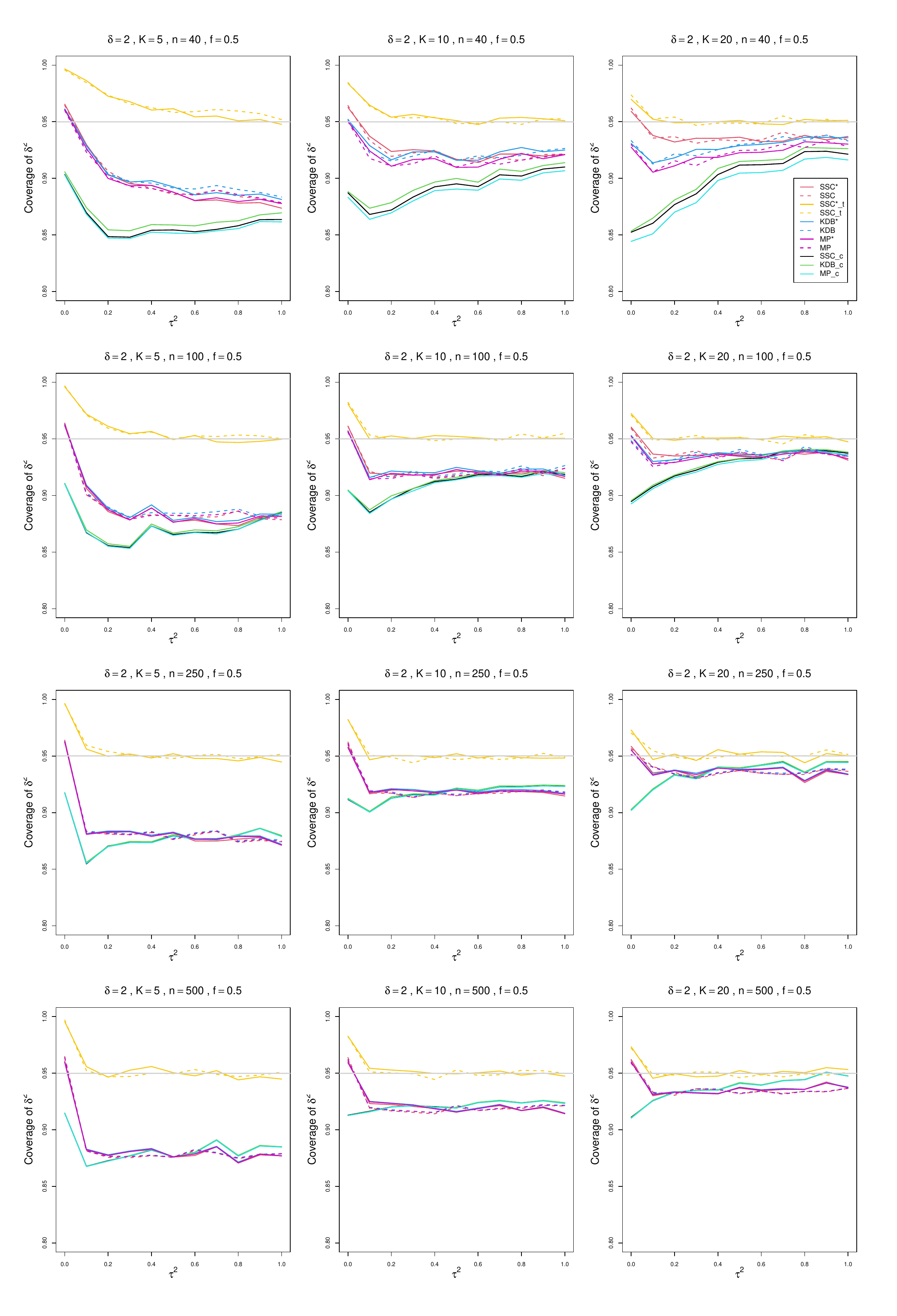}
	\caption{Coverage  of confidence intervals of $\delta^2$ at 95\% nominal level:   MP, KD, SMC and SMC-t intervals, na\"ive and corrected(*),  based on the signed SMD values and conditional intervals  (MP\_c, KD\_c  and  SMC\_c) vs $\tau^2$, for equal sample sizes $n = 40, \;100,\;250,\;500$, $K=5,\;10$ and $20$, $\delta = 2$ and  $f = 0.5$. }
	\label{PlotCovOfDelta2Ksmall_equal_sample_sizes.pdf}
\end{figure}

\begin{figure}[ht]
	\centering
	\includegraphics[scale=0.33]{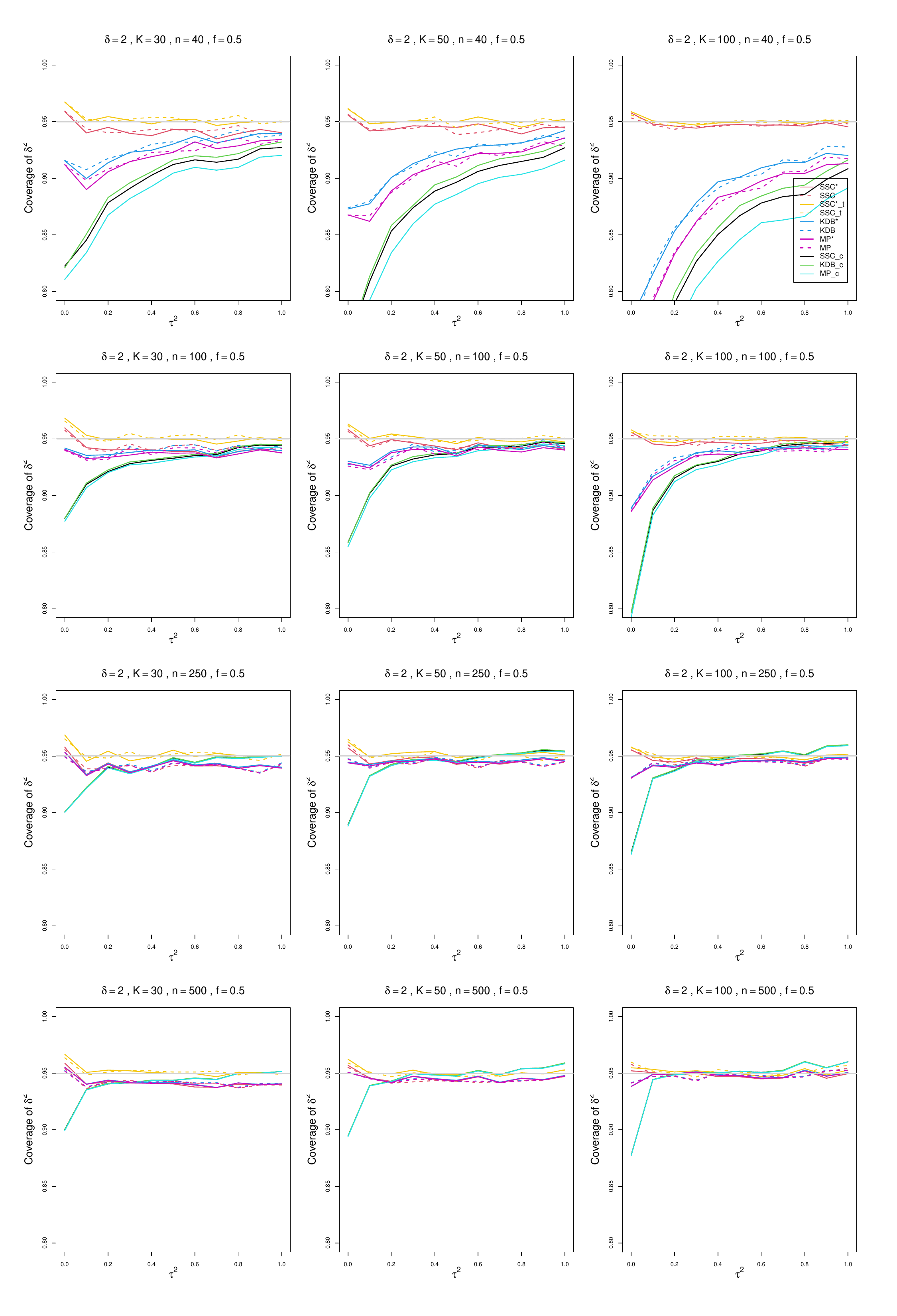}
	\caption{Coverage  of confidence intervals of $\delta^2$ at 95\% nominal level:   MP, KD, SMC and SMC-t intervals, na\"ive and corrected(*),  based on the signed SMD values and conditional intervals  (MP\_c, KD\_c  and  SMC\_c) vs $\tau^2$, for equal sample sizes $n = 100$, $K=30,\;50$ and $100$, $\delta = 2$  and  $f = 0.5$ f }
	\label{PlotCovOfDelta2Kbig_equal_sample_sizes.pdf}
\end{figure}

\begin{figure}[ht]
	\centering
	\includegraphics[scale=0.33]{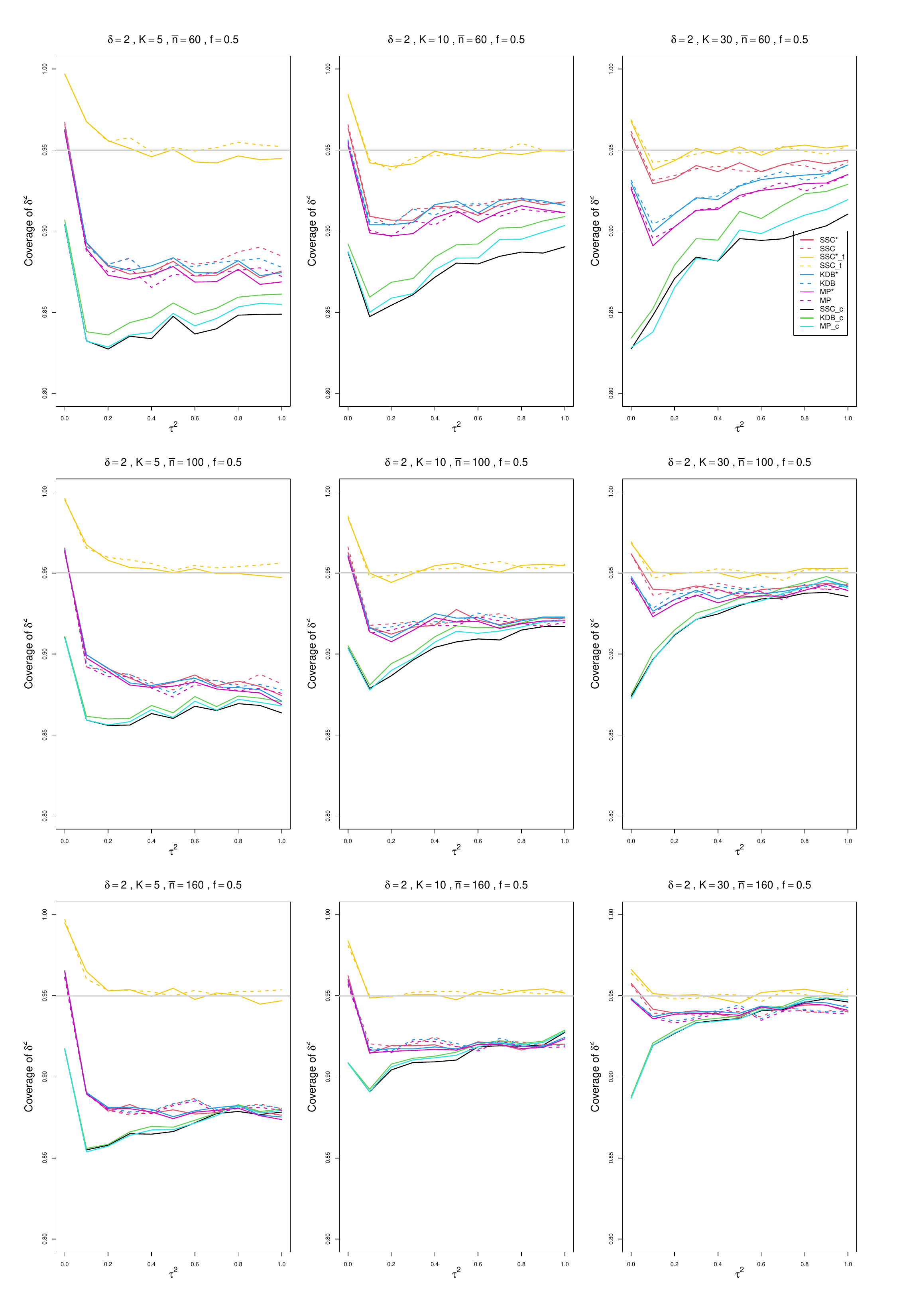}
	\caption{Coverage  of confidence intervals of $\delta^2$ at 95\% nominal level:   MP, KD, SMC and SMC-t intervals, na\"ive and corrected(*),  based on the signed SMD values and conditional intervals  (MP\_c, KD\_c  and  SMC\_c) vs $\tau^2$, for unequal sample sizes $\bar{n} = 60, \;100,\;160$, $K=5,\;10$ and $30$, $\delta = 2$ and  $f = 0.5$. }
	\label{PlotCovOfDelta2Ksmall_unequal_sample_sizes.pdf}
\end{figure}

\end{document}